\documentclass[12pt, draftclsnofoot, onecolumn]{IEEEtran}

\usepackage{graphicx,psfrag,epsfig,epsf,latexsym,hhline,amssymb,multirow}
\usepackage[utf8]{inputenc}
\usepackage[T1]{fontenc}
\usepackage[cmex10]{amsmath}
\usepackage{pst-plot}
\usepackage{pstricks-add}
\usepackage{pifont}
\usepackage{graphicx}
\usepackage{amsthm}
\usepackage{algorithm}
\usepackage{algorithmic}
\usepackage[noadjust]{cite}
\usepackage{array}
\usepackage{blindtext}
\usepackage{etoolbox}
\usepackage{comment}
\usepackage{epstopdf}
\usepackage{hyperref}
\usepackage{bbm}
\usepackage{arydshln}
\usepackage{lipsum}
\newcommand\xqed[1]{%
  \leavevmode\unskip\penalty9999 \hbox{}\nobreak\hfill
  \quad\hbox{#1}}
\newcommand\demo{\xqed{$\blacksquare$}}

\newcolumntype{M}[1]{>{\centering\arraybackslash}m{#1}}
\newcolumntype{P}[1]{>{\centering\arraybackslash}p{#1}}

\newcommand{\iid}{i.\@i.\@d.\ }

\DeclareMathOperator*{\argmin}{arg\,min}

\theoremstyle{definition}
\newtheorem{definition}{Definition}
\newtheorem{example}{Example}
\newtheorem{theorem}{Theorem}
\newtheorem{lemma}{Lemma}
\newtheorem{proposition}{Proposition}
\newtheorem{remark}{Remark}
\newtheorem{corollary}{Corollary}

\begin{document}
\title{Generalized Spatially-Coupled Parallel Concatenated Codes With Partial Repetition}
\author{Min Qiu, Xiaowei Wu, Jinhong Yuan, and Alexandre Graell i Amat

%\thanks{The work of xxx is supported by xxx.}

\thanks{This work was presented in part at the 2021 IEEE Internal Symposium on Information Theory (ISIT) \cite{GSCPCC2021ISIT}.

M. Qiu, X. Wu and J. Yuan are with the School of Electrical Engineering and Telecommunications, University of New South Wales, Sydney, NSW, 2052 Australia (e-mail: min.qiu@unsw.edu.au; xiaowei.wu@unsw.edu.au; j.yuan@unsw.edu.au).

A. Graell i Amat is with the Department of Electrical Engineering, Chalmers University of Technology, SE-41296 Gothenburg, Sweden (e-mail: alexandre.graell@chalmers.se).
}%
}

\maketitle

\begin{abstract}
A new class of spatially-coupled turbo-like codes (SC-TCs), dubbed generalized spatially coupled parallel concatenated codes (GSC-PCCs), is introduced. These codes are constructed by applying spatial coupling on parallel concatenated codes (PCCs) with a fraction of information bits repeated $q$ times. GSC-PCCs can be seen as a generalization of the original spatially-coupled parallel concatenated codes proposed by Moloudi \emph{et al.} \cite{8002601}. To characterize the asymptotic performance of GSC-PCCs, we derive the corresponding density evolution equations and compute their decoding thresholds. The threshold saturation effect is observed and proven. Most importantly, we rigorously prove that any rate-$R$ GSC-PCC ensemble with 2-state convolutional component codes achieves at least a fraction $1-\frac{R}{R+q}$ of the capacity of the binary erasure channel (BEC) for repetition factor $q\geq2$ and this multiplicative gap vanishes as $q$ tends to infinity. To the best of our knowledge, this is the first class of SC-TCs that are proven to be capacity-achieving. Further, the connection between the strength of the component codes, the decoding thresholds of GSC-PCCs, and the repetition factor are established. The superiority of the proposed codes with finite blocklength is exemplified by comparing their error performance with that of existing SC-TCs via computer simulations.
\end{abstract}


\section{Introduction}
Turbo codes \cite{397441,Vucetic:2000:TCP:352869} and low-density parity-check (LDPC) codes \cite{Gallager63low-densityparity-check} are two important classes of codes that have been adopted in various communications standards. These codes are capable of achieving near-Shannon-limit performance as the blocklength grows large. Spatial coupling brings further performance improvement to these codes. The first spatially-coupled LDPC (SC-LDPC) codes, also known as LDPC convolutional codes, were introduced in \cite{782171}. These codes can be obtained by spreading the edges of the Tanner graph \cite{1056404} of the underlying uncoupled LDPC block codes to several adjacent blocks. The most important property of SC-LDPC codes, observed numerically in \cite{5571910} and proven analytically in \cite{5695130,6589171}, is that their iterative decoding threshold under suboptimal belief propagation (BP) decoding achieves the optimal maximum-a-posteriori (MAP) decoding threshold. Such a phenomenon is known as threshold saturation \cite{5695130}. Another advantage of SC-LDPC codes is that they also preserve the minimum distance growth rate of their underlying uncoupled LDPC block codes \cite{7152893}.

The concept of spatial coupling has also been applied, with much success, to various classes of codes to construct capacity-approaching channel codes. For example, the authors in \cite{Smith12} proposed a class of spatially-coupled product codes called staircase codes, which can operate close to the binary symmetric channel capacity under iterative bounded-distance decoding. In this work, we focus on turbo-like codes, whose factor graphs \cite{910572} have convolutional code trellis constraints. In \cite{8002601}, the authors introduced spatially-coupled turbo-like codes (SC-TCs) by applying spatial coupling on parallel concatenated codes (PCCs) \cite{397441}, serially concatenated codes (SCCs) \cite{669119} and braided convolutional codes (BCCs) \cite{5361461}. It was proven in \cite{8002601} that threshold saturation also occurs for SC-TCs. Further investigations on the trade-off between error floor and waterfall performance of SC-TCs as well as the effects of coupling memory and component code blocklength on decoding performance were conducted in \cite{8631116} and \cite{9448689}, respectively. Despite the capacity-approaching performance of SC-SCCs and SC-BCCs, the threshold of SC-PCCs (especially when punctured) are (strictly) bounded away from capacity \cite{8002601}. Recently, partially-information coupled turbo codes (PIC-TCs) were proposed in \cite{8368318} to enhance the performance of the hybrid automatic repeat request protocol in LTE \cite{4907407}. The main idea is that each pair of adjacent code blocks share a fraction of information bits such that these bits are protected by two component turbo codewords. We extended the design of PIC-TCs to a large coupling memory and used density evolution to compute their decoding thresholds in \cite{8989359,PIC2020}. One benefit of such construction is that the technique of partial coupling can be applied to any systematic linear code such as LDPC codes \cite{8301547} and polar codes \cite{9491085} without changing its encoding and decoding architecture. Both theoretical analysis and simulation results in \cite{PIC2020,9174156} showed that partially-coupled turbo codes outperform SC-PCCs and have comparable performance to SC-SCCs and SC-BCCs. However, threshold saturation was neither observed nor proven for PIC-TCs in \cite{8368318,8989359,PIC2020,9174156}.

Although the above works on SC-TCs have all reported capacity-approaching performance, it remains unclear whether spatial coupling can allow turbo-like codes to eventually achieve capacity. Motivated by the fact that PCCs (or turbo codes) are the standard channel coding scheme in the 4G wireless mobile communication systems which coexist with 5G systems, we are interested in designing new and powerful coupled codes with PCCs as component codes that can be compatible with the current standard. In this paper, we introduce generalized SC-PCCs (GSC-PCCs), which are constructed by applying spatial coupling on a component PCC, where a fraction of the information bits are repeated $q$ times. The main contributions of the papers are as follows:
\begin{itemize}
\item We introduce the construction and decoding for GSC-PCCs. We emphasize that the proposed codes not only can be seen as a generalization of the conventional SC-PCCs \cite{8002601}, but also exhibit a similar structure to that of PIC-TCs \cite{PIC2020}, as the repeated bits are protected by the component PCC codewords at several time instants. The proposed construction allows GSC-PCCs to inherit all the positive features of both SC-PCCs and PIC-TCs, such as threshold saturation and close-to-capacity performance when punctured.

\item We derive the density evolution (DE) equations for the proposed GSC-PCC ensembles on the binary erasure channel (BEC). To evaluate and compare the ensembles at rates higher than their mother PCCs, we also derive DE equations for the punctured ensembles. In particular, for a given target code rate $R$ and coupling memory $m$, we find the optimal fraction of repeated information bits that gives the largest decoding threshold for various repetition factors $q$. With these DE equations, we compute the MAP decoding threshold by using the area theorem \cite{Measson2006thesis} and observe threshold saturation numerically.

\item We analytically prove that threshold saturation occurs for the proposed GSC-PCC ensembles by using the proof technique based on potential functions \cite{6325197}. By utilizing this property, we then rigorously prove that the proposed GSC-PCC ensembles with rate $R$ and 2-state convolutional component codes achieve at least a fraction $1-\frac{R}{R+q}$ of the BEC capacity and this multiplicative gap vanishes as $q$ tends to infinity, i.e., GSC-PCCs with 2-state convolutional component codes achieve capacity. To the best of our knowledge, this is the first class of turbo-like codes that are proven to be capacity-achieving. We conjecture that GSC-PCC ensembles with any convolutional component codes are also capable of achieving capacity. Furthermore, the connections between the threshold of GSC-PCC ensembles, the repetition factor, and the strength of the underlying component code are established.

\item The error performance of GSC-PCCs under finite blocklength on the BEC and additive white Gaussian
noise (AWGN) channel is investigated via simulation. Both theoretical analysis and simulation results show that the proposed codes significantly outperform existing coupled codes with PCCs as component codes. In addition, we present an effective method for selecting coupled information bits to further enhance the error performance of GSC-PCCs.
\end{itemize}

%The rest of the paper is organized as follows. Section \ref{sec2} introduces the constructions of the uncoupled PCCs with partial repetition as well as the proposed GSC-PCCs. In Section \ref{sec:DE}, we derive the density evolution equations for both uncoupled and coupled ensembles and determine their iterative decoding thresholds by optimizing the fraction of the repeated information bits. In Section \ref{sec:TS}, we prove threshold saturation analytically for GSC-PCC ensembles and use this result to further prove that any GSC-PCC ensemble with 2-state convolutional component codes achieves capacity. Some useful properties for better understanding and explaining the superior decoding performance of the proposed codes are also revealed and discussed in this section. Numerical results on the error performance under finite blocklength and an effective method for choosing the coupling bits are presented in Section \ref{sec:sim}. Finally, the paper is concluded in Section \ref{sec:conclude}.

%\subsection{Notation}
%Scalars and vectors are written in lightface and boldface letters, respectively, e.g., $x$ and $\mathbf{x}$.

\section{Generalized Spatially-Coupled Parallel Concatenated Codes}\label{sec2}
In this section, we first introduce the uncoupled PCCs with partial information repetition that will be used to construct GSC-PCCs. Then, we present the encoding and decoding of GSC-PCCs.

\subsection{Parallel Concatenated Codes with Partial Repetition}
Uncoupled PCCs with partial information repetition are similar to the dual-repeat-punctured turbo codes in \cite{5308225}, except that in our case only a fraction of the information bits are repeated. The encoder of an uncoupled PCC with partial repetition is depicted in Fig. \ref{fig:GSC_PCC_ENC}(a). A length-$K$ information sequence $\boldsymbol{u}$ is divided into two sequences, $\boldsymbol{u}_{\text{r}}$ and $\boldsymbol{u}_{\text{o}}$. Then, sequence $\boldsymbol{u}_{\text{r}}$ is repeated $q$ times and combined with $\boldsymbol{u}_{\text{o}}$ to form a length-$K'$ information sequence $[\boldsymbol{u}_{\text{r}},\ldots,\boldsymbol{u}_{\text{r}},\boldsymbol{u}_{\text{o}}]$. The resultant sequence and its reordered copy $\Pi([\boldsymbol{u}_{\text{r}},\ldots,\boldsymbol{u}_{\text{r}},\boldsymbol{u}_{\text{o}}])$, where $\Pi(.)$ denotes the interleaving function, are encoded by the upper and lower convolutional encoders, respectively. We define the repetition ratio $\lambda \triangleq \frac{K'-K}{(q-1)K'} \in [0,1/q]$ as the length of $\boldsymbol{u}_r$ over $K'$. The length of $\boldsymbol{u}_{\text{o}}$ is then given by $(1-q\lambda)K'$. The repetition ratio is an important parameter and its definition and notation are used throughout the rest of the paper. Finally, the codeword is a length-$N $ sequence $\boldsymbol{c}=[\boldsymbol{u}_{\text{r}},\boldsymbol{u}_{\text{o}},\boldsymbol{v}^{\text{U}},\boldsymbol{v}^{\text{L}}]=[\boldsymbol{u},\boldsymbol{v}^{\text{U}},\boldsymbol{v}^{\text{L}}]$, comprising the information sequence before repetition, as well as two length-$\frac{N-K}{2}$ parity sequences generated by the upper and lower convolutional encoders. Note that it is natural to exclude all other $q-1$ replicas of $\boldsymbol{u}_{\text{r}}$ from $\boldsymbol{c}$ as they do not contain new information. Given the code rate of the mother PCC, $R_0 = \frac{K'}{N'}$, where $N' = N-K+K'$ is its codeword length, the code rate of the uncoupled PCC with partial repetition is %as
\begin{align}\label{uncoupled_rate}
R_{\text{uc}}  =& \frac{K'(1-(q-1)\lambda)}{N'-K'+K'(1-(q-1)\lambda)}  %\\
= \frac{1-(q-1)\lambda}{\frac{1}{R_0}-(q-1)\lambda}  %\\
\geq \frac{1}{q(\frac{1}{R_0}-1)+1},
\end{align}
where the last inequality shows that the lowest rate is achieved when $\lambda = 1/q$.

\subsection{Encoding}
%In this section, we show the construction of GSC-PCCs with partial repetition PCCs as the component codes. Note that the notations defined above are also used in this section.

\begin{figure}[t!]
%\begin{minipage}[b]{0.49\linewidth}
	\centering
\includegraphics[width=2.7in,clip,keepaspectratio]{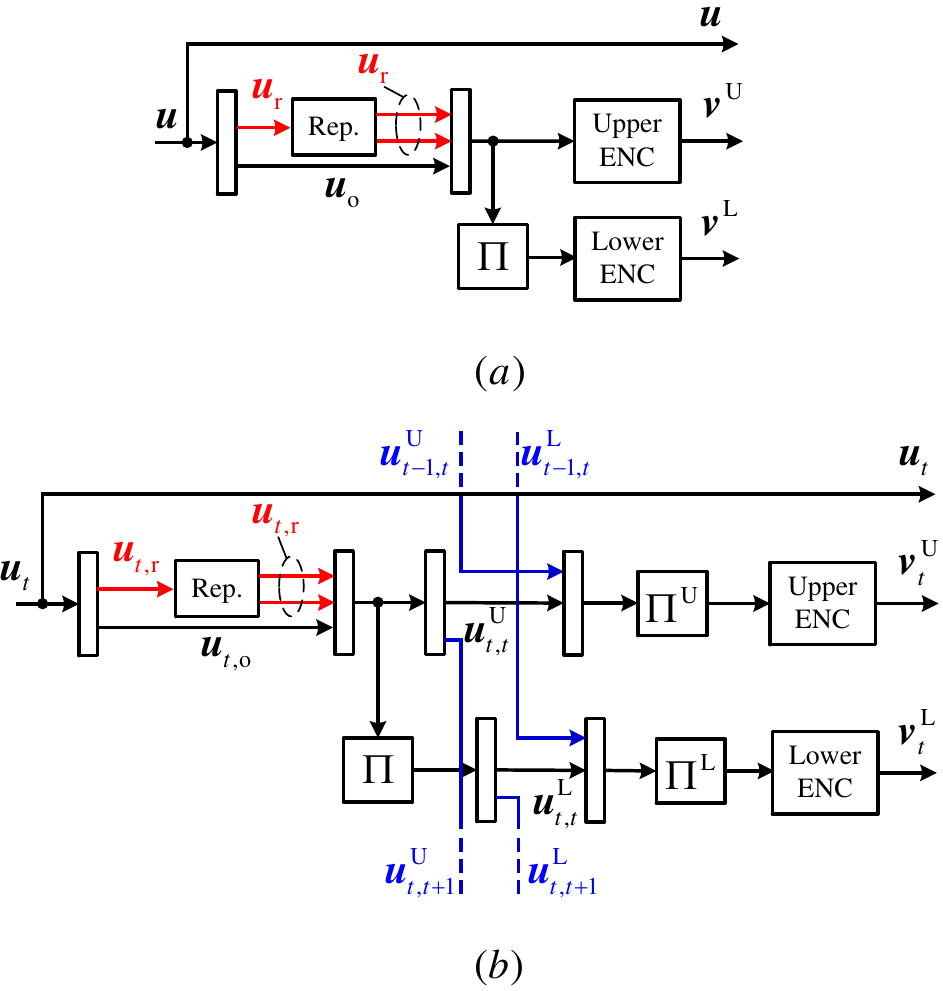}
\caption{Encoders of (a) an uncoupled PCC with partial repetition, and (b) a GSC-PCC with $m=1$ at time $t$.}
\label{fig:GSC_PCC_ENC}
\end{figure}
%\end{minipage}\hfill
%\begin{minipage}[b]{0.49\linewidth}

We construct GSC-PCCs by applying spatial coupling to the above PCCs with partial repetition. The block diagram of a GSC-PCC with coupling memory $m=1$ at time instant $t$ is depicted in Fig. \ref{fig:GSC_PCC_ENC}(b).

An information sequence $\boldsymbol{u}$ is divided into $L$ sequences of equal length $K$, which are denoted by $\boldsymbol{u}_t$, $t =1,\ldots,L$. We refer to $L$ as the coupling length. At time $t$, $\boldsymbol{u}_t$ is decomposed into $\boldsymbol{u}_{t,\text{r}}$ and $\boldsymbol{u}_{t,\text{o}}$, where $\boldsymbol{u}_{t,\text{r}}$ is a length-$\lambda K'$ sequence and $\boldsymbol{u}_{t,\text{o}}$ is a length-$K'(1-q\lambda)$ sequence. Sequence $\boldsymbol{u}_{t,\text{r}}$ is repeated $q$ times and combined with $\boldsymbol{u}_{t,\text{o}}$ to form a length-$K'$ information sequence $[\boldsymbol{u}_{t,\text{r}},\ldots,\boldsymbol{u}_{t,\text{r}},\boldsymbol{u}_{t,\text{o}}]$. The resultant sequence is then decomposed into $m+1$ sequences of length $\frac{K'}{m+1}$, denoted by $\boldsymbol{u}^\text{U}_{t,t+j}$, $j=0,\ldots,m$. The information sequence $\boldsymbol{u}^\text{U}_{t,t+j}$ is used as a part of the input of the upper convolutional encoder at time $t+j$. The coupling is performed such that a length-$K'$ information sequence, $[\boldsymbol{u}^\text{U}_{t-m,t},\ldots,\boldsymbol{u}^\text{U}_{t,t}]$, is formed. Meanwhile, the reordered copy of information sequence $[\boldsymbol{u}_{t,\text{r}},\ldots,\boldsymbol{u}_{t,\text{r}},\boldsymbol{u}_{t,\text{o}}]$, i.e., $\Pi([\boldsymbol{u}_{t,\text{r}},\ldots,\boldsymbol{u}_{t,\text{r}},\boldsymbol{u}_{t,\text{o}}])$, is also decomposed into $m+1$ sequences of length $\frac{K'}{m+1}$, i.e., $\boldsymbol{u}^\text{L}_{t,t+j}$, $j=0,\ldots,m$, where $\boldsymbol{u}^\text{L}_{t,t+j}$ is used as a part of the input of the lower convolutional encoder at time $t+j$. With coupling, a length-$K'$ information sequence $[\boldsymbol{u}^\text{L}_{t-m,t},\ldots,\boldsymbol{u}^\text{L}_{t,t}]$ is formed. The codeword obtained at time $t$ is a length-$N$ sequence $\boldsymbol{c}_t = [\boldsymbol{u}_t,\boldsymbol{v}^\text{U}_t,\boldsymbol{v}^\text{L}_t]$, where $\boldsymbol{v}^\text{U}_t$ and $\boldsymbol{v}^\text{L}_t$ are two length-$\frac{N-K}{2}$ parity sequences as the result of encoding $\Pi^\text{U}([\boldsymbol{u}^\text{U}_{t-m,t},\ldots,\boldsymbol{u}^\text{U}_{t,t}])$ and $\Pi^\text{L}([\boldsymbol{u}^\text{L}_{t-m,t},\ldots,\boldsymbol{u}^\text{L}_{t,t}])$ at the upper and lower systematic convolutional encoders, respectively, at time $t$. We remark that the three interleavers are crucial for introducing randomness in code structures such that the codes become ensembles for which density evolution can be rigorously applied to analyze the decoding threshold. %Notice that the upper and lower interleavers, $\Pi^\text{U}$ and $\Pi^\text{L}$, are used together with $\Pi$, as shown in Fig. \ref{fig:GSC_PCC_ENC}(b), to ensure that the minimum distance of GSC-PCCs is not smaller than that of the underlying uncoupled PCC codes. This is based on a similar argument for SC-PCCs in \cite[Theorem 1]{8631116}.

To initialize and terminate the coupled chain, we can simply set $\boldsymbol{u}_t$ to $\boldsymbol{0}$ for $t\leq0$ and $t>L$. As a result, the code rate of the GSC-PCC with coupling memory $m$, coupling length $L$, and repetition factor $q$ is
\begin{align}
R_{\text{sc}}&=\frac{KL}{NL+m(N-K)}%\nonumber \\
=\frac{K'L(1-(q-1)\lambda)}{L(N'-K'(q-1)\lambda)+m(N'-K')} \nonumber \\
&=\frac{L(1-(q-1)\lambda)}{L(\frac{1}{R_0}-(q-1)\lambda)+m(\frac{1}{R_0}-1)},
\end{align}
where $R_0=\frac{K'}{N'}$ is the code rate of the mother PCC. When $L  \rightarrow \infty$, the code rate of the GSC-PCC approaches $R_{\text{uc}}$ in \eqref{uncoupled_rate}.

Due to partial repetition of information bits, GSC-PCCs have an encoding latency of $\frac{1}{1-(q-1)\lambda}$ times higher than that of SC-PCCs. When $q=2$, GSC-PCCs have a similar encoding latency to PIC-TCs because PIC-TCs also have a fraction of information bits repeated twice. However, it is important to note that the encoding of GSC-PCCs can be performed either in parallel, i.e., encoding $L$ information sequences in parallel, or in a serial and streaming fashion, making them still more appealing than block codes.

\subsection{Comparison to Existing Codes}
There are connections between the proposed GSC-PCCs and some existing SC-TCs. First, the proposed codes can be seen as a generalization of the conventional SC-PCCs \cite{8002601}. More precisely, one can obtain the original SC-PCC from a GSC-PCC by setting either $q=1$ or $\lambda = 0$. However, the introduction of partial repetition gives rise to a significant performance improvement, as it will be shown in Section \ref{sec:DE} and Section \ref{sec:TS}. In addition, the proposed codes have more flexible structure as they can reach a code rate as low as $\frac{1}{2q+1}$ when the mother PCC is rate-$1/3$, while the lowest code rate for the conventional SC-PCCs is $1/3$.

The proposed codes also bear similarities to PIC-TCs \cite{PIC2020}, whose coupled information bits are encoded (and protected) by two turbo encoders (four convolutional encoders). In fact, PIC-TCs can be seen as having a fraction of information bits repeated twice and using the copies of those information bits as the input of the turbo encoder at the succeeding time instant. For GSC-PCCs, this can happen when some of the information bits from $\boldsymbol{u}_{t,\text{r}}$ appear in $\boldsymbol{u}^\text{U}_{t,t}$ and $\boldsymbol{u}^\text{L}_{t,t}$, while their copies appear in $\boldsymbol{u}^\text{U}_{t,t+j}$ and $\boldsymbol{u}^\text{L}_{t,t+j}$, $j\in\{1,\ldots,m\}$. In this case, those repeated and coupled information bits can be protected by the component PCC codewords at multiple time instants. However, the coupling of PIC-TCs is at the turbo code level (the information encoded by upper and lower encoders is the same). In contrast, GSC-PCCs are coupled at the convolutional code level (the information encoded by upper and lower encoders is different) such that the proposed codes inherit many nice properties from SC-PCCs, such as threshold saturation (crucial for achieving capacity) and decoding threshold improvement from employing stronger convolutional component codes.

\subsection{Decoding}
The decoding of GSC-PCCs consists of two types of iterations: intra-block iterations and inter-block iterations. Specifically, an intra-block iteration is the exchange of the extrinsic information of information bits between the upper and lower Bahl–Cocke–Jelinek–Raviv (BCJR) \cite{1055186} component decoders at the same time instant. An inter-block iteration is the exchange of extrinsic information of information bits in component codes across $L$ time instants in a forward/backward round trip. Note that the inter-block iteration can also be performed in a sliding window fashion with window size $W$, where $m+1 \leq W \leq L $. To avoid repetition, we focus on the log-likelihood ratio (LLR) updates for information bits since the LLR updates for all parity bits are the same as those for the conventional uncoupled turbo codes.

Let $u^{\text{U}}_{t,k}$ denote the $k$-th information bit at the upper decoder at time $t$ and $k \in \{1,\ldots,K'\}$. Let $L_{\text{C}}(.)$, $L_{\text{E}}(.)$ denote the channel and extrinsic LLRs, respectively. In addition, we denote by $\mathcal{Q}^{\text{U}}_k$ the sets of bit positions associated with $u^{\text{U}}_{t,k}$ and its replicas which appear in the upper decoder at the same instant, where $\mathcal{Q}^{\text{U}}_k\subseteq \{1,\ldots,K'\}$ and $|\mathcal{Q}^{\text{U}}_k|\subseteq \{1,\ldots,q\}$. The definition of $\mathcal{Q}^{\text{L}}_k$ is analogous to $\mathcal{Q}^{\text{U}}_k$. Notice that $t$ is not required here because the three interleavers and the selection of information bits to be repeated or coupled are the same for all $t\in\{1,\ldots,L\}$. As an example, $|\mathcal{Q}^{\text{U}}_k|=1$ means that either $u^{\text{U}}_{t,k}$ is not repeated or its replicas are not in the upper decoder at the same time instant. Consequently, there are two types of \emph{a priori} LLRs of each repeated information bit: the \emph{a priori} LLR obtained from its replicas at the upper BCJR decoder at the same time instant, denoted by $L_{\text{A}_1}(.)$; and the \emph{a priori} LLR obtained from its interleaved bit at the lower BCJR decoder at the same time instant, denoted by $L_{\text{A}_2}(.)$. Furthermore, we define $L_{\text{in}}(.)$ and $L_{\text{out}}(.)$ as the input and output LLR of the BCJR decoder. Due to space limitations, we omit the LLR updates for inter-block decoding as it is similar to SC-PCCs \cite{8002601}. The updates for the LLR associated with information bits during intra-block decoding are described as follows.

\emph{\textbf{Step 1 (Input LLR Computation):}} Construct the LLR of $u^{\text{U}}_{t,k}$ for the upper BCJR decoder input as $L_{\text{in}}(u^{\text{U}}_{t,k}) = L_{\text{C}}(u^{\text{U}}_{t,k})+L_{\text{A}_1}(u^{\text{U}}_{t,k})+L_{\text{A}_2}(u^{\text{U}}_{t,k})$, where $L_{\text{A}_1}(u^{\text{U}}_{t,k})$ is computed in Step 4 in the last iteration, and $L_{\text{A}_2}(u^{\text{U}}_{t,k})$ is obtained from the extrinsic LLR of its interleaved bit $u^{\text{L}}_{t,\tilde{k}}$ at the lower BCJR component decoder with a step analogous to Step 4.

\emph{\textbf{Step 2 (BCJR Component Decoding):}} Perform BCJR decoding and obtain the output LLR of $u^{\text{U}}_{t,k}$ as $L_{\text{out}}(u^{\text{U}}_{t,k})$.

\emph{\textbf{Step 3 (Extrinsic Information Computation):}} The extrinsic LLR of $u^{\text{U}}_{t,k}$ is computed as $L_{\text{E}}(u^{\text{U}}_{t,k}) = \sum_{k \in \mathcal{Q}^{\text{U}}_k} \hat{L}_{\text{E}}(u^{\text{U}}_{t,k})$, where we define $\hat{L}_{\text{E}}(u^{\text{U}}_{t,k})\triangleq L_{\text{out}}(u^{\text{U}}_{t,k})-L_{\text{in}}(u^{\text{U}}_{t,k})$. Then, for any $k,k'\in \mathcal{Q}^{\text{U}}_k$ and $k\neq k'$, we have $L_{\text{E}}(u^{\text{U}}_{t,k}) =L_{\text{E}}(u^{\text{U}}_{t,k'})$.

\emph{\textbf{Step 4 (A priori Information Computation):}} Compute the \emph{a priori} LLR of $u^{\text{U}}_{t,k}$ to be used in the next iteration as $L_{\text{A}_1}(u^{\text{U}}_{t,k})= L_{\text{E}}(u^{\text{U}}_{t,k})-\hat{L}_{\text{E}}(u^{\text{U}}_{t,k})$. The extrinsic LLR of $u^{\text{U}}_{t,k}$ is used as the \emph{a priori} LLR of its interleaved bit $u^{\text{L}}_{t,\tilde{k}}$ at the lower decoder, i.e., $L_{\text{A}_2}(u^{\text{L}}_{t,\tilde{k}}) =  L_{\text{E}}(u^{\text{U}}_{t,k})$. For any $\tilde{k},\tilde{k'} \in \mathcal{Q}^{\text{L}}_{\tilde{k}}$ and $\tilde{k}\neq \tilde{k'}$, $L_{\text{A}_2}(u^{\text{L}}_{t,\tilde{k}}) =L_{\text{A}_2}(u^{\text{L}}_{t,\tilde{k'}})$.

After Step 4, the intra-block decoding proceeds to the lower BCJR component decoding, for which the LLR updates can be easily obtained from the above steps by interchanging subscripts U and L. Compared to SC-PCCs, the increase in the complexity mainly comes from Step 3, where additional computation resource is required to perform the combination of the extrinsic information associated with the repeated bits. Moreover, addition memory is needed to store the bit positions of repeated bits (does not change with $t$). However, when $q=2$, the complexity of GSC-PCCs is comparable to that of PIC-TCs since PIC-TCs also have a fraction of information bits repeated twice.

\section{Density Evolution Analysis on the BEC}\label{sec:DE}
In this section, we first look into the graph representation of GSC-PCCs and then derive the exact density evolution equations to characterize their decoding threshold. In this work, we consider a rate $R_0 = 1/3$ mother PCC built from two rate-$1/2$ recursive systematic convolutional codes.

\subsection{Graph Representation}
Turbo-like code ensembles can be represented by a compact graph \cite{8002601}, which simplifies the factor graph representation. The main idea is that each information or parity sequence in a factor graph is represented by a single variable node, while a trellis constraint is represented by a factor node. An interleaver is represented by a line segment that crosses an edge.

\begin{figure}[t!]
	\centering
\includegraphics[width=2.4in,clip,keepaspectratio]{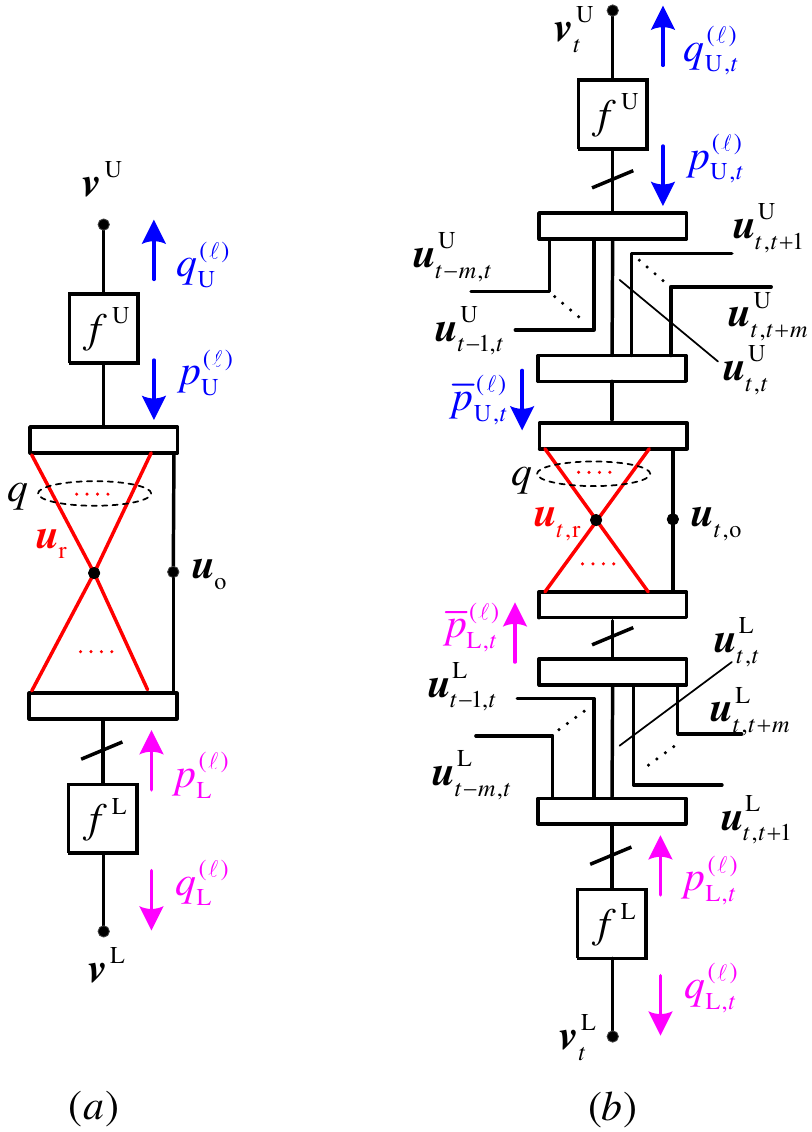}
%\vspace{-6mm}
\caption{Compact graph representation of (a) uncoupled ensembles, and (b) GSC-PCC ensembles at time $t$.}
\label{fig:uc_graph}
%\end{minipage}\hfill
%\vspace{-6mm}
\end{figure}

We first look at the compact graph of an uncoupled PCC with partial repetition, which is depicted in Fig. \ref{fig:uc_graph}(a). Compared to the compact graph of a conventional PCC (see \cite[Fig. 4a]{8002601}), the difference is that in our case the information node $\boldsymbol{u}$ is represented by two nodes, $\boldsymbol{u}_{\text{r}}$ and $\boldsymbol{u}_{\text{o}}$.\footnote{With some abuse of language, we sometimes refer to a variable node
representing a sequence as the sequence itself.} Since the information sequence $\boldsymbol{u}_{\text{r}}$ is repeated $q$ times before being encoded by the PCC encoder, node $\boldsymbol{u}_{\text{r}}$ connects the upper and lower factor nodes $f^\text{U}$ and $f^\text{L}$ via $q$ edges, respectively.

The compact graph representation of GSC-PCCs with coupling memory $m$ and at time $t$ is depicted in Fig. \ref{fig:uc_graph}(b). It is similar to the compact graph of SC-PCCs (see \cite[Fig. 5a]{8002601}), except that information node $\boldsymbol{u}_t$ is represented by nodes $\boldsymbol{u}_{t,\text{r}}$ and $\boldsymbol{u}_{t,\text{o}}$, where node $\boldsymbol{u}_{t,\text{r}}$ connects the upper and lower factor nodes via $q$ edges, respectively. Analogous to many turbo-like ensembles, as the lengths of each component codeword and random interleavers go to infinity, the assumptions of decoder symmetry, all-one codeword, concentration, asymptotically tree-like computation graph for a fixed number of iterations can be adopted to the graphs of GSC-PCC ensembles \cite[Ch. 6]{Richardson:2008:MCT:1795974}. This allows us to rigorously apply DE to analyze the decoding threshold of GSC-PCC ensembles.
%In the next section, we derive the DE equations for the BEC based on the graphs in Fig. \ref{fig:uc_graph}.

\subsection{Density Evolution}
Since GSC-PCCs are newly proposed, it is natural to study their behavior under a fundamental channel model, i.e., the BEC model, first. For this model, the exact decoding threshold for turbo-like codes can be rigorously analyzed \cite{8002601}. In addition, the results in \cite{6589171,8631116,PIC2020} suggest that several good classes of spatially-coupled codes for the BEC also perform well over other channels. Hence, we focus on the BEC in this work in order to fully understand the behavior of the proposed codes.

Let $\epsilon$ denote the channel erasure probability of the BEC. For a rate-$1/2$ convolutional code, we let $f^\text{U}_\text{s}(.)$ and $f^\text{U}_\text{p}(.)$ denote the transfer functions of the upper decoder for information and parity bits, respectively, where $x$ and $y$ correspond to the input erasure probabilities for information and parity bits, respectively. Similarly, let $f^\text{L}_\text{s}(.)$ and $f^\text{L}_\text{p}(.)$ denote the transfer functions of the lower decoder for information and parity bits, respectively. The exact input/output transfer functions of a convolutional code under the BCJR decoding \cite{1055186} on the BEC can be explicitly derived by using the methods in \cite{370145,1258535}.

\subsubsection{Uncoupled Ensembles}\label{sec:un_de}
As shown in Fig. \ref{fig:uc_graph}(a), $p^{(\ell)}_{\text{U}}$ and $q^{(\ell)}_{\text{U}}$ represent the output erasure probability of factor node $f^{\text{U}}$ for information and parity bits, respectively, after $\ell$ decoding iterations. Similarly, $p^{(\ell)}_{\text{L}}$ and $q^{(\ell)}_{\text{L}}$ denote the output erasure probability of $f^{\text{L}}$ for information and parity bits, respectively.

The DE update equation for the output erasure probability of the information bits at node $f^\text{U}$ is
\begin{align}\label{eq:un_de1}
p^{(\ell)}_{\text{U}}
 =f^\text{U}_{\text{s}}\left(\epsilon q\lambda \left(p^{(\ell-1)}_{\text{U}}\right)^{q-1} \left(p^{(\ell)}_\text{L}\right)^q + \epsilon\left(1-q\lambda\right)p^{(\ell)}_{\text{L}},\epsilon\right),
%q^{(\ell)}_{\text{U}}
%=f^\text{U}_{\text{p}}\left(\epsilon q\lambda \left(p^{(\ell-1)}_{\text{U}}\right)^{q-1} \left(p^{(\ell)}_\text{L}\right)^q+\epsilon\left(1-q\lambda\right)p^{(\ell)}_{\text{L}},\epsilon\right),\label{eq:un_de2}
\end{align}
where $(1-q\lambda)$ and $q\lambda$ are the weights of the erasure probability of $\boldsymbol{u}_{\text{o}}$ and $\boldsymbol{u}_{\text{r}}$, respectively, determined by the ratios of their lengths over $K'$ (the input length of the upper and lower convolutional encoder), $\epsilon q\lambda(p^{(\ell-1)}_{\text{U}})^{q-1} (p^{(\ell)}_\text{L})^q$ is the weighted extrinsic erasure probability from node $\boldsymbol{u}_{\text{r}}$ to node $f^\text{U}$ while the powers on $p^{(\ell-1)}_{\text{U}}$ and $p^{(\ell)}_\text{L}$ are due to the repetition at the upper and lower encoders, $\epsilon(1-q\lambda)p^{(\ell)}_{\text{L}}$ is the weighted extrinsic erasure probability from node $\boldsymbol{u}_{\text{o}}$ to node $f^\text{U}$, and finally the average erasure probability from node $\boldsymbol{v}^\text{U}$ to node $f^\text{U}$ is $\epsilon$.

The DE update equation for the output erasure probability of the parity bits at node $f^\text{U}$, i.e., $q^{(\ell)}_{\text{U}}$, can be obtained by replacing the transfer function $f^\text{U}_{\text{s}}(.)$ by $f^\text{U}_{\text{p}}(.)$. To obtain the DE update equations for $p^{(\ell)}_{\text{L}}$ and $q^{(\ell)}_{\text{L}}$ at node $f^\text{L}$, we can simply interchange $p_{\text{U}}$ and $p_{\text{L}}$ and replace $f^\text{U}(.)$ by $f^\text{L}(.)$ in \eqref{eq:un_de1}.

\subsubsection{Coupled Ensembles}
Based on the compact graph in Fig. \ref{fig:uc_graph}(b), we denote by $p^{(\ell)}_{\text{U},t}$ and $q^{(\ell)}_{\text{U},t}$ the output erasure probability of $f^{\text{U}}$ for information and parity bits, respectively, at time $t$ and after $\ell$ decoding iterations. Similarly, $p^{(\ell)}_{\text{L},t}$ and $q^{(\ell)}_{\text{L},t}$ denote the output erasure probability of $f^{\text{L}}$ for information and parity bits, respectively. We also define the average erasure probability from $f^{\text{U}}$ and $f^{\text{L}}$ to $\boldsymbol{u}_t$ as $\bar{p}^{(\ell-1)}_{\text{U},t}$ and $\bar{p}^{(\ell-1)}_{\text{L},t}$, respectively, where
\begin{align}
\bar{p}^{(\ell-1)}_{\text{U},t} = \frac{1}{m+1}\sum_{j=0}^m p^{(\ell-1)}_{\text{U},t+j},\label{eq:sc_de_1}\\
\bar{p}^{(\ell-1)}_{\text{L},t} = \frac{1}{m+1}\sum_{j=0}^m p^{(\ell-1)}_{\text{L},t+j}.\label{eq:sc_de_2}
\end{align}

By using \eqref{eq:sc_de_1} and \eqref{eq:sc_de_2}, as well as taking into account the partial repetition of information bits, we obtain the DE update for the erasure probability of the information bits at $f^{\text{U}}$ as
\begin{align}
p^{(\ell)}_{\text{U},t}
=&f^\text{U}_\text{s}\Bigg(\frac{\epsilon}{m+1}\sum_{k=0}^m \Big(q\lambda \left(\bar{p}^{(\ell-1)}_{\text{L},t-k}\right)^q \left(\bar{p}^{(\ell-1)}_{\text{U},t-k}\right)^{q-1}
+\left(1-q\lambda\right)\bar{p}^{(\ell-1)}_{\text{L},t-k}\Big),\epsilon  \Bigg) \nonumber \\
 = &f^\text{U}_\text{s}\Bigg(\frac{\epsilon}{m+1}\sum_{k=0}^m \Bigg(q\lambda \left(\frac{1}{m+1}\sum_{j=0}^m p^{(\ell-1)}_{\text{L},t+j-k}\right)^q
 \cdot\left(\frac{1}{m+1}\sum_{j=0}^m p^{(\ell-1)}_{\text{U},t+j-k}\right)^{q-1} \nonumber \\
 &+\frac{1-q\lambda}{m+1}\sum_{j=0}^m p^{(\ell-1)}_{\text{L},t+j-k}\Bigg),\epsilon  \Bigg). \label{eq:sc_de_3}
\end{align}

To avoid repetition, we omit the DE equations for the erasure probability of the parity bits at $f^{\text{U}}$ as well as the DE equations at node $f^\text{L}$ as they can be trivially obtained from \eqref{eq:sc_de_3}.

\subsection{Random Puncturing}
To increase the code rate, we consider random puncturing of parity bits.

Let $\rho\in [0,1]$ denote the fraction of surviving parity bits after puncturing. For such a randomly punctured code sequence transmitted over the BEC with erasure probability $\epsilon$, the erasure probability of the parity sequence becomes $\epsilon_{\rho} = 1-(1-\epsilon)\rho$ \cite[Eq. 4]{7353121}. As a result, the DE equations for the punctured uncoupled and coupled ensembles can be obtained by substituting $\epsilon_{\rho} \rightarrow \epsilon$ for the average erasure probability from node $\boldsymbol{v}^\text{U}$ to node $f^\text{U}$ in \eqref{eq:un_de1} and that from node $\boldsymbol{v}_t^\text{U}$ to node $f^\text{U}$ in \eqref{eq:sc_de_3}, respectively.

After puncturing, the code rates of both uncoupled and coupled ensembles (considering $L\rightarrow \infty$) become
\begin{align}\label{eq:rate_punc}
R =\frac{1-(q-1)\lambda}{\left(\frac{1}{R_0}-1\right)\rho+1-(q-1)\lambda}.
\end{align}
Given $(R_0,R,q,\lambda)$, then $\rho$ is uniquely determined.

\subsection{Decoding Thresholds}
We compute the decoding thresholds over the BEC by using the DE equations derived in the previous section. We consider 4-state, rate-$1/2$ convolutional encoders with generator polynomial $(1,5/7)$ in octal notation for both upper and lower encoders. Given a target code rate $R \in \left[\frac{1}{q\left(\frac{1}{R_0}-1\right)+1},1\right)$ and coupling memory $m$, we optimize the repetition ratio $\lambda$ in order to maximize the iterative decoding threshold for various $q$. The optimized repetition ratios and the corresponding thresholds for the uncoupled ensembles (denoted by $\lambda$ and $\epsilon_{\text{BP}}$, respectively)\footnote{The decoding of turbo-like codes comprises BCJR decoding for convolutional component codes while the message exchange between BCJR component decoders follows the extrinsic message passing rule. Hence, we refer to the threshold under iterative message passing decoding with BCJR component decoding as BP threshold.} and coupled ensembles with coupling memory $m$ (denoted by $\lambda^{(m)}$ and $\epsilon^{(m)}_{\text{BP}}$, respectively) are reported in Table \ref{table0} and Table \ref{table1}, respectively. Note that the optimal $\lambda$ could be a range of values because these $\lambda$ lead to the same decoding threshold up to the fourth decimal place, which we believe have sufficient accuracy. In Table \ref{table1}, we also report the MAP threshold of the uncoupled ensembles ($\epsilon_{\text{MAP}}$), the minimum coupling memory ($m_{\min}$) for which threshold saturation is observed numerically, and the gap between the decoding threshold $\epsilon^{(m=m_{\min})}_{\text{BP}}$ and the corresponding BEC capacity (denoted by $\Delta_{\text{SH}}=1-R-\epsilon^{(m=m_{\min})}_{\text{BP}}$). Since turbo-like code ensembles including uncoupled PCCs with partial repetition, can be described by using factor graphs, the MAP threshold can be computed by using the area theorem \footnote{Although the MAP threshold given by the area theorem is an upper bound, we opt to drop the term ``upper bound'' for simplicity as the numerical results show that the thresholds of the coupled ensembles converge to this upper bound.} \cite{Measson2006thesis}
\begin{align}\label{eq:MAP_find}
R = \int_{\epsilon_{\text{MAP}}}^1 h^{\text{BP}}(\epsilon) d \epsilon \overset{(a)}{=} \int_{\epsilon_{\text{MAP}}}^1 R \bar{p}(\epsilon)+(1-R)\bar{q}(\epsilon) d \epsilon,
\end{align}
where $R$ is the target code rate, $h^{\text{BP}}(\epsilon)$ is the BP extrinsic information transfer (EXIT) function, $\bar{p}(\epsilon)$ and $\bar{q}(\epsilon)$ are the average extrinsic erasure probability for information bits and parity bits, respectively, $(a)$ follows from \cite{1523540}. To be specific,
\begin{align}
\bar{p}(\epsilon) =&q\lambda \left(p^{(\infty)}_{\text{U}}\right)^{q} \left(p^{(\infty)}_\text{L}\right)^q + \left(1-q\lambda\right)p^{(\infty)}_{\text{U}}p^{(\infty)}_{\text{L}},  \\
\bar{q}(\epsilon) =& \frac{1}{2}f^\text{U}_{\text{p}}\left(\epsilon q\lambda \left(p^{(\infty)}_{\text{U}}\right)^{q-1} \left(p^{(\infty)}_\text{L}\right)^q + \epsilon\left(1-q\lambda\right)p^{(\infty)}_{\text{L}},1-(1-\epsilon)\rho\right) \nonumber \\
&+\frac{1}{2}f^\text{L}_{\text{p}}\left(\epsilon q\lambda \left(p^{(\infty)}_{\text{L}}\right)^{q-1} \left(p^{(\infty)}_\text{U}\right)^q + \epsilon\left(1-q\lambda\right)p^{(\infty)}_{\text{U}},1-(1-\epsilon)\rho\right).
\end{align}

\begin{table}[t!]
%\begin{minipage}[b]{0.47\linewidth}
%\tiny
%\setlength\tabcolsep{3pt}
  \centering
 \caption{Optimal Repetition Ratio of GSC-PCCs}\label{table0}
\begin{tabular}{c c  c  c  c  c }
\hline
 Rate   & $q$ & $\lambda$  &  $\lambda^{(m=1)}$   & $\lambda^{(m=3)}$  & $\lambda^{(m=5)}$  \\
 \hline
$3/4$  &  2 &$[0.287,0.313]$  & 0.5    & 0.5 & 0.5 \\
$3/4$  &  4 &0.172  & $[0.201,0.206]$    &  0.24  &  0.25 \\
$3/4$  &  6 &0.13  &   0.137   &  $[0.152,0.154]$  &   $[0.162,0.163]$ \\
\hline
 $1/2$  & 2  & $[0.184,0.213]$    & 0.44  &  0.5 & 0.5  \\
 $1/2$  & 4  &  0.147    & $[0.187,0.188]$  & 0.23  &  0.25  \\
  $1/2$  & 6  &  0.12    & 0.131   &  $[0.150,0.151]$ &  $[0.156,0.160]$  \\
\hline
$1/3$  &  2 &$[0.088,0.124]$  & $[0.37,0.39]$   & 0.5  & 0.5 \\
$1/3$  &  4 & $[0.107,0.108]$  & $[0.162,0.172]$   & $[0.216,0.229]$   & 0.25  \\
$1/3$  &  6 & $[0.104,0.105]$  &  $[0.121,0.122]$  &   $[0.138,0.146]$ & $[0.151,0.158]$  \\
  \hline
  $1/4$  &  2 &$[0.036,0.072]$  & $[0.319,0.353]$   & 0.5  & 0.5 \\
$1/4$  &  4 & $[0.083,0.086]$  & $[0.152,0.162]$   & $[0.216,0.229]$   & 0.24  \\
$1/4$  &  6 & $0.112$  &  $[0.112,0.116]$  &   $[0.134,0.143]$ & $[0.143,0.158]$  \\
  \hline
\end{tabular}
%\vspace{-6mm}
\end{table}
%\end{minipage}\hfill
%\begin{minipage}[b]{0.51\linewidth}
\begin{table}[t!]
%\setlength\tabcolsep{0.9pt}
%\tiny
%\setlength\tabcolsep{3pt}
  \centering
 \caption{Iterative Decoding thresholds of GSC-PCCs, SC-PCCs and PIC-TCs}\label{table1}
 %\vspace{-3mm}
\begin{tabular}{c  c  c  c  c  c  c  c  c  c}
\hline
 Rate & Ensemble & $q$ & $\epsilon_{\text{BP}}$  &  $\epsilon^{(m=1)}_{\text{BP}}$   & $\epsilon^{(m=3)}_{\text{BP}}$  & $\epsilon^{(m=5)}_{\text{BP}}$& $\epsilon_{\text{MAP}}$ &$m_{\min}$ & $\Delta_{\text{SH}}$ \\
 \hline
 &PIC-TC &  2 &-  & 0.2307    & 0.2337 & 0.2344& 0.2351 & 1000 & 0.0149\\
 \cdashline{2-10}
  &SC-PCC &  1 &0.1854  & 0.1876    & 0.1876  & 0.1876 & 0.1876  & 1& 0.0624\\
   \cdashline{2-10}
$3/4$ &   &  2 &0.2115  & 0.2326    & 0.2352 & 0.2352& 0.2352 & 3 & 0.0148\\
   &GSC-PCC &  4 &0.2268  & 0.2380    & 0.2430  & 0.2443  &0.2444 & 6 & 0.0056\\
 &  &  6 &0.2218  &  0.2406    &  0.2442 &  0.2457 & 0.2466 & 9 & 0.0034\\
\hline
 &PIC-TC &  2 &-  & 0.4865    & 0.4906 & 0.4920& 0.4934 &1000 & 0.0066\\
 \cdashline{2-10}
  &SC-PCC &  1 &0.4606  & 0.4689    & 0.4689  & 0.4689 & 0.4689  & 1 & 0.0311\\
   \cdashline{2-10}
 $1/2$  & & 2  & 0.4698    & 0.4907  &   0.4938 & 0.4938 & 0.4938 & 3 & 0.0062\\
 &GSC-PCC & 4  &   0.4849   & 0.4940  &  0.4969   &  0.4978  & 0.4979 & 6 & 0.0021\\
     & & 6  &   0.4747   &  0.4952  &  0.4974   &  0.4982 & 0.4988 & 9 & 0.0012 \\
\hline
 &PIC-TC &  2 &-  & 0.6576    & 0.6615 & 0.6625& 0.6640 & 1000 & 0.0027\\
 \cdashline{2-10}
  &SC-PCC &  1 &0.6428  & 0.6553    & 0.6553  & 0.6553 & 0.6553  & 1 & 0.0113\\
   \cdashline{2-10}
$1/3$   & &  2 & 0.6446   & 0.6627   & 0.6647  & 0.6647 & 0.6647 & 3 & 0.0020\\
&GSC-PCC &  4 &  0.6583   & 0.6642   &  0.6656  &  0.6660 & 0.6661 & 6 & 0.0006\\
  & &  6 &    0.6512  &   0.6648  &   0.6658 &   0.6661& 0.6663 & 8 & 0.0004\\
  \hline
   &PIC-TC &  2 &-  & 0.7425    & 0.7459 & 0.7466& 0.7483 & 1000 & 0.0017\\
 \cdashline{2-10}
 \multirow{2}{*}{$1/4$} & &  2 & 0.7313   & 0.7478   & 0.7491  & 0.7491 & 0.7491 & 3 & 0.0009\\
& GSC-PCC&  4 &  0.7413   & 0.7487   &  0.7495  &  0.7497 & 0.7497  & 5 & 0.0003\\
  & &  6 &    0.7406  &   0.7490  &   0.7496 &   0.7497 & 0.7498 & 6 & 0.0002\\
  \hline
\end{tabular}
%\end{minipage}\hfill
%\vspace{-6mm}
\end{table}

For comparison purposes, we list the decoding thresholds of SC-PCCs \cite{8002601} and PIC-TCs \cite{PIC2020}, which all use the same convolutional encoder as that for GSC-PCCs, in Table \ref{table1}. Except the rate-$1/3$ SC-PCC which reaches its lowest rate, the rest of the codes all require puncturing on the parity bits. Note that SC-PCCs can be seen as a special case of the proposed GSC-PCCs with $q=1$ or $\lambda = 0$ while PIC-TCs only have a fraction of information bits repeated twice, i.e., $q=2$. Since PIC-TCs do not show threshold saturation \cite{PIC2020}, their MAP threshold and the BP threshold of the underlying uncoupled ensemble are unknown. Hence, we show their iterative decoding threshold for $m=1000$ under the column of $\epsilon_{\text{MAP}}$.

First, it can be observed that the thresholds of GSC-PCCs surpass those of PIC-TCs and SC-PCCs for the same coupling memories and same code rates even for $q=2$ and puncturing. Although all codes exhibit a larger $\Delta_{\text{SH}}$ with increasing the fraction of punctured bits, the proposed GSC-PCCs can close this gap by increasing $q$. Particularly, the BP thresholds of GSC-PCCs improve with increasing $q$ for all the considered code rates and coupling memories. On the other hand, uncoupled PCCs with partial repetition have worse performance than coupled ensembles and their BP thresholds do not always improve with $q$. Intuitively, the BP threshold would improve if the extrinsic information of each convolutional decoder, i.e., BCJR decoder, becomes more reliable. However, increasing the repetition factor $q$ does not necessarily lead to more reliable extrinsic information as the large number of repetition (without coupling) could cause some bias. Meanwhile, puncturing is required to compensate the code rate reduction introduced by repetition, which subsequently reduces the BP threshold. For a large enough coupling memory, threshold saturation effect can be observed for GSC-PCCs. It is also worth noting that the optimal repetition ratio $\lambda$ approaches $1/q$ when $m$ is large in most cases, e.g., $m\geq m_{\min}$, suggesting that choosing $\lambda=1/q$ is sufficient for the proposed codes to universally achieve their MAP thresholds.

\begin{figure}[t!]
%\begin{minipage}[b]{0.49\linewidth}
	\centering
\includegraphics[width=3.1in,clip,keepaspectratio]{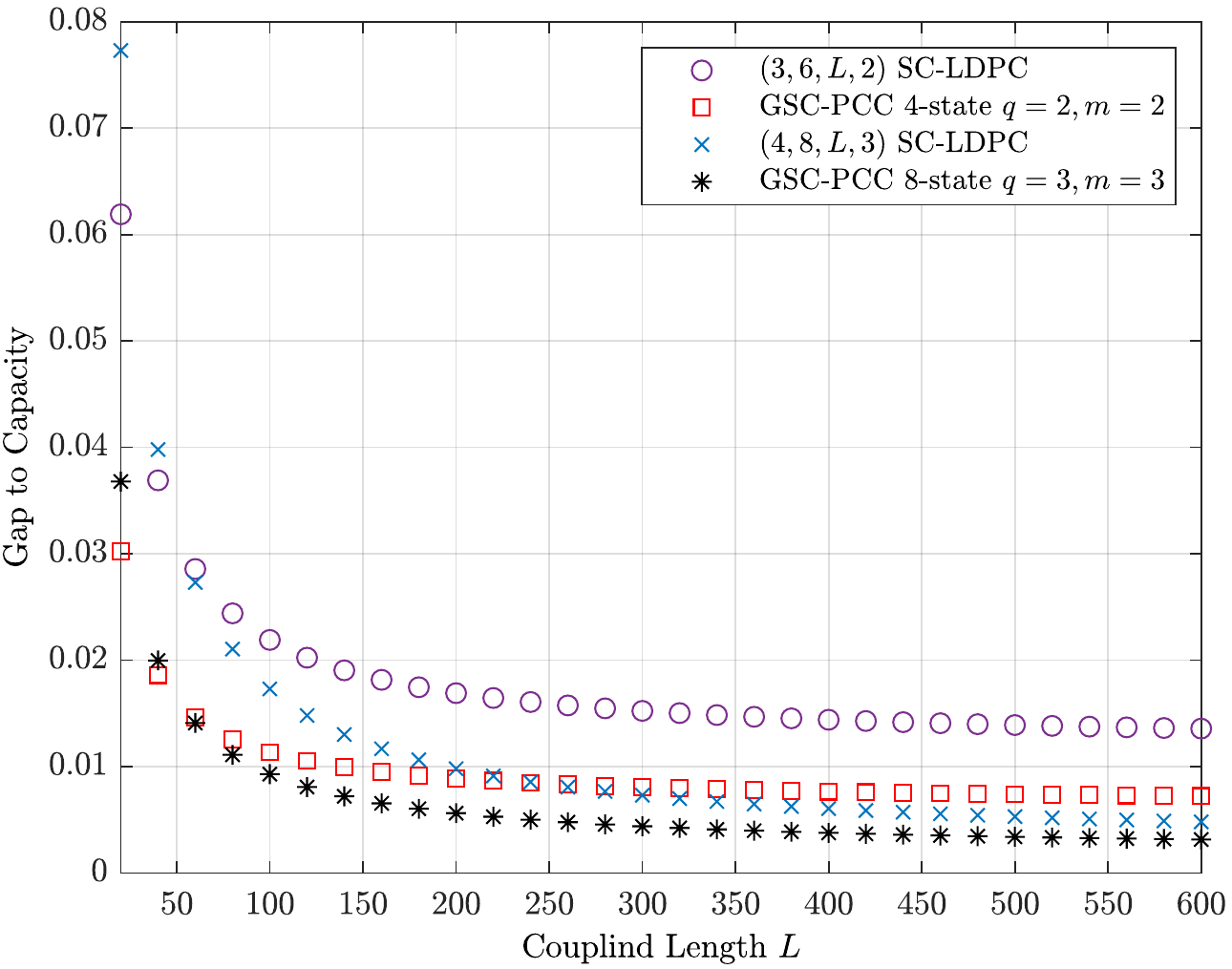}
%\vspace{-6mm}
\caption{Gap to the BEC capacity for GSC-PCCs and SC-LDPC codes with target rate $1/2$.}
\label{fig:gap_compare}
\end{figure}

We also compare the decoding thresholds between GSC-PCCs and SC-LDPC codes \cite{7152893}. Since both SC-LDPC codes and GSC-PCCs are capacity-achieving (as we will see in Section \ref{sec:cap_achieve}), we are interested in their performance by taking into account rate loss due to termination, i.e., under finite coupling length $L$. As an example, we consider two GSC-PCC ensembles with $(1,5/7)$ and $(1,15/13)$ convolutional component codes with $q\in\{2,3\}$, $\lambda \in \{0.5,0.3\}$, and $m \in \{2,3\}$, respectively. Moreover, we consider a $(3,6)$ SC-LDPC ensemble and a $(4,8)$ SC-LDPC ensemble with coupling widths 2 and 3, respectively, as two benchmark codes. We denote by $R_{\text{term}}$ the design rate of a terminated spatially coupled ensemble. The gaps to the BEC capacity ($1-R_{\text{term}}-\epsilon^{(m)}_{\text{BP}}$) versus $L$ for the aforementioned four codes are shown in Fig. \ref{fig:gap_compare}.

Observe that the proposed GSC-PCC ensembles have a smaller gap to capacity than that for the SC-LDPC ensembles with the same coupling memory (width). This is because GSC-PCC ensembles have less rate loss and a larger threshold than SC-LDPC ensembles. For example, the GSC-PCC ensemble with $(q,L,m)=(2,50,2)$ has a rate 0.4950 and a threshold of 0.4936 while the $(3,6,50,2)$ SC-LDPC ensemble has a rate of 0.48 and a threshold of 0.4881. Hence, the proposed GSC-PCCs have rate and threshold advantages over SC-LDPC codes.

\section{Threshold Saturation And Capacity-Achieving}\label{sec:TS}
In this section, we first analytically prove that threshold saturation occurs for GSC-PCCs. We then utilize this property to further prove that the proposed codes achieve capacity. Finally, some useful properties in relation to the threshold behavior of GSC-PCCs are presented.

\subsection{Threshold Saturation}
We consider identical upper and lower encoders for simplicity. Thus, for uncoupled PCCs with partial repetition, we can define $f_\text{s}\triangleq f^\text{U}_\text{s} = f^\text{L}_\text{s}$ and $x^{(\ell)}\triangleq p^{(\ell)}_{\text{L}} = p^{(\ell)}_\text{U}$. The DE equation in \eqref{eq:un_de1} can be written as a fixed point recursive equation
\begin{subequations}\label{eq:pot1}
\begin{align}
x^{(\ell)}= &f_\text{s}\bigg(q\epsilon\lambda \left(x^{(\ell-1)}\right)^{2q-1} +\epsilon(1-q\lambda)x^{(\ell-1)},1-(1-\epsilon)\rho \bigg)\label{eq:pot1a} \\
=&f\left( g\left(x^{(\ell-1)}\right);\epsilon \right),\label{eq:pot1b}
\end{align}
\end{subequations}
where $\eqref{eq:pot1b}$ is due to using the following definitions
\begin{align}
f(x ; \epsilon) &\triangleq f_\text{s}(\epsilon x, 1-(1-\epsilon)\rho), \label{eq:f}\\
g(x)&\triangleq q\lambda x^{2q-1}+(1-q\lambda)x. \label{eq:g}
\end{align}
First, we note that the following properties hold due to \cite[Lemma 1]{8002601} and \cite[Lemma 2]{8002601}:

1) $f(x ; \epsilon)$ is increasing in both arguments $x,\epsilon \in(0,1]$;

2) $f(0;\epsilon)=f(\epsilon;0)=g(0)=0$;

3) $f(x ; \epsilon)$ has continuous second derivatives on $[0,1]$ with respect to all arguments.

Moreover, it is easy to see that $g'(x)>0,\forall x \in (0,1]$, and $g''(x)$ exists and is continuous $\forall x \in[0,1]$. Therefore, the DE recursion in \eqref{eq:pot1} forms a scalar admissible system \cite[Def. 1]{6325197}.

For the above scalar admissible system, the potential function \cite[Def. 2]{6325197} is
\begin{subequations}\label{eq:pot2}
\begin{align}
U(x;\epsilon) =& xg(x)-G(x)-F(g(x);\epsilon) \label{eq:pot2a} \\
=& \left(q-\frac{1}{2}\right)\lambda x^{2q}+\frac{1}{2}(1-q\lambda)x^2 %\nonumber \\
- \int_0^{q\lambda x^{2q-1}+(1-q\lambda)x} f_\text{s}(\epsilon z,1-(1-\epsilon)\rho)dz,\label{eq:pot2b}
%& = (q-\frac{1}{2})1/q x^{2q}- \int_0^{ x^{2q-1}} F_s(\epsilon z,1-(1-\epsilon)\rho)dz \\
%& =  (1- \frac{1}{2q})x^{2q} - \int_0^{ x^{2q-1}} F_s(\epsilon z,1-(1-\epsilon)\rho)dz
\end{align}
\end{subequations}
where $\eqref{eq:pot2b}$ follows from
\begin{align}
F(x;\epsilon) &= \int_0^x f(z;\epsilon) dz = \int_0^x f_\text{s}(\epsilon z, 1-(1-\epsilon)\rho) dz, \\
G(x) &= \int_0^x g(z) dz = \frac{1}{2}\lambda x^{2q}+\frac{1}{2}(1-q\lambda)x^2.
\end{align}

The following definitions are useful in the subsequent analysis.
\begin{definition}\label{def:sst}
The single system threshold of an admissible system is defined as \cite{6325197,6887298}
\begin{align}
\epsilon_\text{s} = \sup\left\{\epsilon \in [0,1]:U'(x;\epsilon)>0,\forall x \in (0,1] \right\}.
\end{align}
\end{definition}
In our case, $\epsilon_\text{s}$ is the BP threshold of the uncoupled ensembles. The fixed point for the recursive equation in \eqref{eq:pot1} is $x=0$ for $\epsilon <\epsilon_\text{s}$, and converges to a non-zero fixed point otherwise.
\begin{definition}\label{def:2}
The potential threshold of an admissible system is defined as \cite{6325197,6887298}
\begin{align}
\epsilon_\text{c} = \sup\left\{\epsilon \in [0,1]:\min_{x\in[u(\epsilon),1]}U(x;\epsilon)\geq 0,u(\epsilon)>0 \right\},
\end{align}
where
\begin{align}
u(\epsilon) = \sup\left\{\tilde{x}\in[0,1]: f( g(x);\epsilon )<x,x\in (0,\tilde{x})\right\},
\end{align}
is the minimum unstable fixed point for $\epsilon>\epsilon_\text{s}$.
\end{definition}

\begin{figure}[t!]
	\centering
\includegraphics[width=3.1in,clip,keepaspectratio]{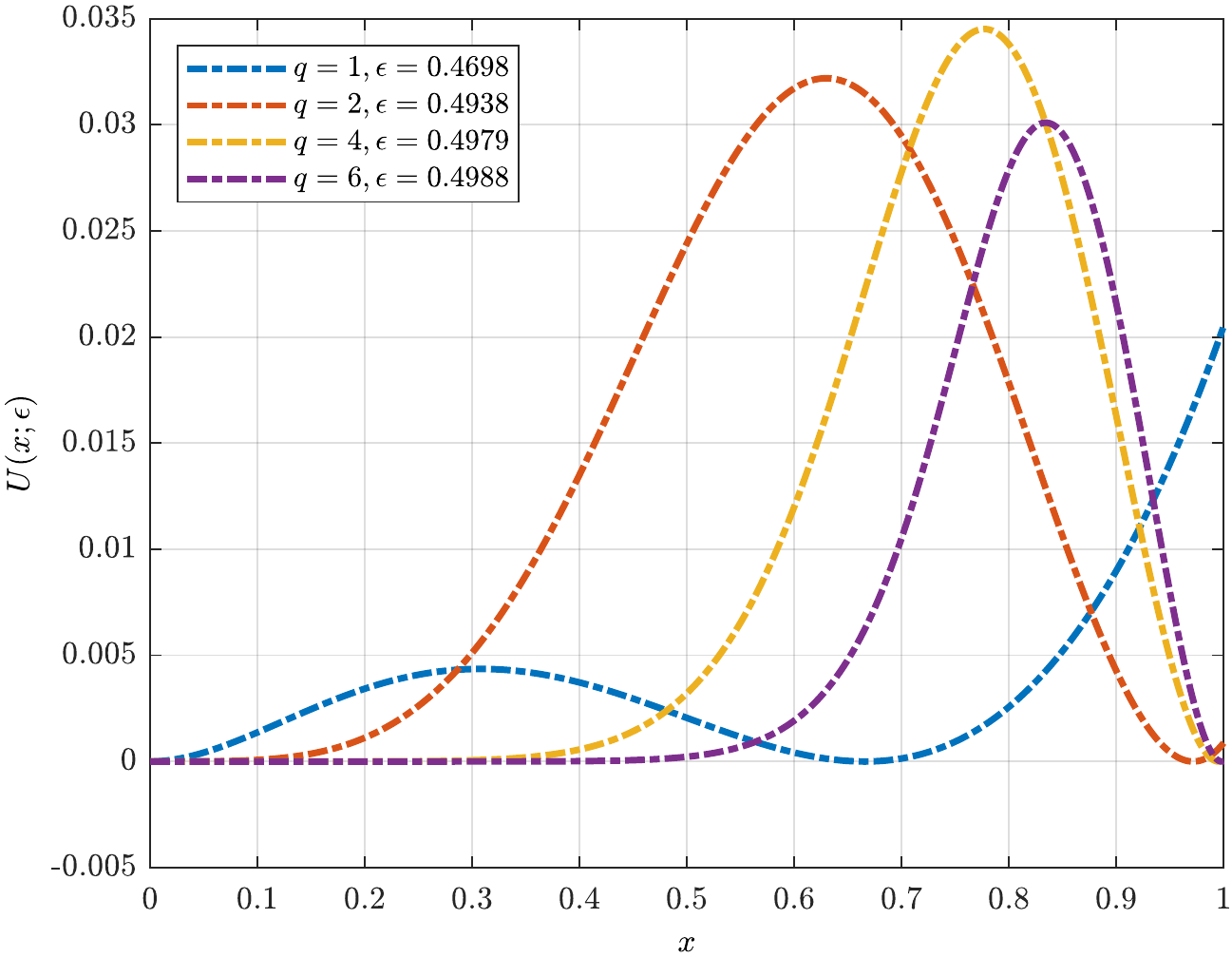}
%\vspace{-6mm}
\caption{Potential functions of the uncoupled PCC ensembles with $\lambda = 1/q$ for rate-$1/2$.}
\label{fig:potential_fun_2}
%\end{minipage}\hfill
%\vspace{-6mm}
\end{figure}

\begin{example}\label{example1}
The potential functions for the rate-$1/2$ uncoupled ensemble built from two $(1,5/7)$ convolutional codes for various $q$ are shown in Fig. \ref{fig:potential_fun_2}. In this example, we set $\lambda = 1/q$. The channel erasure probability $\epsilon$ is set to the values of the potential thresholds, which are shown in the legend of Fig. \ref{fig:potential_fun_2}. It can be seen that the potential thresholds match with the MAP thresholds in Table \ref{table1}. \demo
\end{example}

%\end{minipage}\hfill
%\begin{minipage}[b]{0.49\linewidth}

As for the coupled system, we can rewrite the DE equation from \eqref{eq:sc_de_3} into the following by letting $x^{(\ell)}_t\triangleq \bar{p}^{(\ell)}_{\text{L},t} = \bar{p}^{(\ell)}_{\text{U},t}$.
\begin{subequations}\label{eq:pot3}
\begin{align}
x_t^{(\ell)} =& \frac{1}{1+m}\sum_{j=0}^mf_\text{s}\Bigg( \frac{\epsilon}{1+m}\sum_{k=0}^m \bigg(q\lambda \left(x_{t+j-k}^{(\ell-1)}\right)^{2q-1}
+(1-q\lambda)x_{t+j-k}^{(\ell-1)}\bigg) , 1-(1-\epsilon)\rho \Bigg) \\
=&\frac{1}{1+m}\sum_{j=0}^mf\left( \frac{1}{1+m}\sum_{k=0}^m g\left(x_{t+j-k}^{(\ell-1)}\right); \epsilon \right).
\end{align}
\end{subequations}
Then, we have the following theorem.
\begin{theorem}\label{the:ts}
For the spatially-coupled system defined in \eqref{eq:pot3} and any $\epsilon<\epsilon_\text{c}$, where $\epsilon_\text{c}$ is the potential threshold associated with the potential function in \eqref{eq:pot2}, the only fixed point of the recursion in \eqref{eq:pot3} is $\boldsymbol{x} = \boldsymbol{0}$ as $L \rightarrow \infty$, $m \rightarrow \infty$ and $L \gg m$.
\end{theorem}
\begin{IEEEproof}
The proof follows from \cite[Theorem 1]{6325197}.
\end{IEEEproof}
Therefore, threshold saturation occurs for the proposed GSC-PCC ensembles. As a result, the BP thresholds of GSC-PCCs even when $q$ is very large can be easily found via computing either the potential thresholds by using Definition \ref{def:2} or the MAP thresholds by using the area theorem \cite{1523540} as in \eqref{eq:MAP_find}. Consider GSC-PCCs with identical upper and lower 2-state, 4-state and 8-state component convolutional encoders with generator polynomials $(1,1/3)$, $(1,5/7)$ and $(1,15/13)$, respectively. We report the potential thresholds of the uncoupled ensembles with various $q$ (denoted by $\epsilon^{(q)}_{\text{c}}$) for different rates in Table \ref{MAP1}. Here, we choose $\lambda = 1/q$ as we observe from Tables \ref{table0}-\ref{table1} that this choice allows GSC-PCCs to achieve their respective MAP thresholds as $m$ goes large.

\begin{table}[t!]
  \centering
  %\tiny
 \caption{Potential thresholds of Uncoupled PCCs with Partial Repetition}\label{MAP1}
 %\vspace{-3mm}
\begin{tabular}{c c c c c c c c c c}
\hline
 Rate  & States &   $\epsilon^{(q=1)}_{\text{c}}$ &   $\epsilon^{(q=2)}_{\text{c}}$  & $\epsilon^{(q=3)}_{\text{c}}$ & $\epsilon^{(q=4)}_{\text{c}}$ & $\epsilon^{(q=5)}_{\text{c}}$ & $\epsilon^{(q=6)}_{\text{c}}$ & $\epsilon^{(q=50)}_{\text{c}}$ \\  \hline
 	& 2 	&0.0285 &  0.0751  & 0.0846 & 0.0888& 0.0913& 0.0928 & 0.0992\\
$9/10$	& 4 &	0.0582 &  0.0882  & 0.0932 & 0.0952& 0.0963& 0.0970 & 0.0996\\
 	& 8 &	0.0769&  0.0940  & 0.0966 & 0.0977& 0.0982&  0.0986& 0.0998\\
\hline
  	& 2 &0.0661	 &  0.1582  & 0.1747 & 0.1819& 0.1859& 0.1884 & 0.1987\\
 $4/5$	& 4 &	0.1391 &  0.1848  & 0.1915 & 0.1941& 0.1955& 0.1964 & 0.1996\\
  	& 8 &	0.1698 &  0.1930  & 0.1962 & 0.1975& 0.1981& 0.1985 & 0.1998\\
\hline
   & 2   & 0.0895 & 0.2027  & 0.2217 & 0.2298 & 0.2343& 0.2372 & 0.2486\\
 $3/4$ & 4 &  0.1876  & 0.2352  & 0.2418 & 0.2444 & 0.2457& 0.2466 & 0.2496\\
  & 8  & 0.2204  & 0.2435  & 0.2466 & 0.2477 & 0.2483& 0.2486 & 0.2498\\
\hline
   & 2  & 0.1375  & 0.2811  & 0.3027 & 0.3116 & 0.3165& 0.3196 & 0.3318\\
 $2/3$ & 4 &  0.2772  & 0.3209  & 0.3266 & 0.3288 & 0.3299& 0.3306 & 0.3330\\
   & 8  & 0.3080  & 0.3282  & 0.3307 & 0.3316 & 0.3321& 0.3323 & 0.3332\\
\hline
  & 2  & 0.2808 & 0.4520  & 0.4727 & 0.4809 & 0.4854  &0.4881  & 0.4987\\
 $1/2$ & 4 & 0.4689 & 0.4938  & 0.4968 & 0.4979 & 0.4985  & 0.4988 & 0.4998\\
  & 8  & 0.4863 & 0.4976  & 0.4989 & 0.4993 & 0.4995  & 0.4996 & 0.4999\\
\hline
   & 2  & 0.5000 & 0.6352  & 0.6493 & 0.6548 & 0.6576  &  0.6594 &0.6659\\
  $1/3$ & 4 &  0.6553 & 0.6647  & 0.6657 & 0.6661 & 0.6662  & 0.6663  &0.6667\\
     & 8 & 0.6621 & 0.6659  & 0.6663 & 0.6665 & 0.6665  & 0.6665  &0.6667\\
  \hline
\end{tabular}
%\vspace{-6mm}
\end{table}

Table \ref{MAP1} shows that the potential thresholds of uncoupled PCCs with partial repetition improve as $q$ increases. The thresholds also improve as the number of states of the component convolutional codes increases. When $q$ is large, the potential thresholds of all the ensembles approach the BEC capacity for all the considered rates. In particular, even the potential thresholds for the ensembles with 2-state component convolutional codes are within 0.002 to the BEC capacity when $q =50$. This suggests that the BP thresholds of GSC-PCCs can achieve the BEC capacity as $q$ tends to infinity regardless of the number of states of the component convolutional codes. Hence, one can simply increase the repetition factor $q$ to obtain a GSC-PCC with its decoding threshold very close to the BEC capacity for any given component convolutional code while it is difficult for the SC-TCs in \cite{8002601} to further improve their thresholds without changing the component codes. In the next section, we prove that the proposed GSC-PCCs can in fact achieve the BEC capacity.

\subsection{Achieving Capacity}\label{sec:cap_achieve}
First, we let $\lambda = 1/q$ as this simple choice suffices to allow GSC-PCCs to achieve the largest threshold as $m$ becomes large. As a result, the potential function in \eqref{eq:pot2} simplifies to
\begin{align}\label{eq:pot_final1}
U(x;\epsilon)= \left(1-\frac{1}{2q}\right) x^{2q}- \int_0^{x^{2q-1}} f_\text{s}(\epsilon z,1-(1-\epsilon)\rho)dz,
\end{align}
where $\rho = \frac{R_0(1-R)}{qR(1-R_0)}$ due to \eqref{eq:rate_punc}. Then, we state the main result of this section in the following.

\begin{theorem}\label{the:cap}
The rate-$R$ GSC-PCC ensemble with $(1,1/3)$ convolutional component codes achieves at least a fraction $1-\frac{R}{R+q}$ of the BEC capacity under BP decoding.
\end{theorem}
\begin{IEEEproof}
See Appendix \ref{app:cap}.
\end{IEEEproof}

Corollary \ref{corollary1} follows immediately from Theorem \ref{the:cap}.
\begin{corollary}\label{corollary1}
The GSC-PCC ensemble with $(1,1/3)$ convolutional component codes achieves the BEC capacity under BP decoding as $q \rightarrow \infty$.
\end{corollary}

%With Theorem \ref{the:cap} and Corollary \ref{corollary1}, we therefore conjecture that GSC-PCCs ensembles achieve the BEC capacity as $q \rightarrow \infty$ in general.
%\begin{conjecture}\label{cojecture1}
%GSC-PCCs ensembles achieve the BEC capacity under BP decoding as $q \rightarrow \infty$.
%\end{conjecture}

\begin{remark}
To prove Theorem \ref{the:cap}, we choose to use the potential function as the key tool rather than the area theorem because the potential function only involves the transfer function of the information bits of the component decoder while the area theorem requires the transfer functions of both information and parity bits. It is also interesting to see that the GSC-PCC ensemble constructed from 2-state convolutional component codes has a multiplicative gap to the BEC capacity and the gap vanishes as $q \rightarrow \infty$. Generalizing the result of Theorem \ref{the:cap} to the GSC-PCC ensembles with any component convolutional codes is highly non-trivial because the transfer functions of different component decoders have to be derived separately. In particular, when the number of states is large, the derivation for the transfer function becomes extremely cumbersome and the exact analytical expression would be much more complicated than that of the 2-state code in \eqref{eq:2state} (e.g., \cite[Tables I-II]{1258535}). However, Theorem \ref{the:cap} together with the results of Table \ref{MAP1} strongly suggest that the proposed code ensembles with any given component convolutional codes also achieve capacity.
\end{remark}

Although obtaining an analytical expression for the potential threshold of GSC-PCCs with any given component convolutional codes is difficult, we establish in the next section some useful properties of the proposed codes to allow us to better understand how their decoding thresholds behave.

\subsection{Useful Properties of GSC-PCCs}
In this section, we further investigate some properties of GSC-PCCs by establishing the links between the decoding thresholds of the proposed coupled codes, the strength of the component codes, and the repetition factor (Propositions \ref{lem2}-\ref{lem3} below). Following from the previous analysis, we fix $\lambda = 1/q$.

Since the subsequent analysis only involves the transfer function of the information bits, we simply drop the subscript ``s'' from the transfer function for simplicity. Before proceeding, we present a useful result from \cite[Lemma 7.5]{Measson2006thesis}.
\begin{lemma}
Consider a convolutional code $\mathcal{C}$ with code rate $R_\mathcal{C}\geq 1/2$. Its decoder's transfer function for the information bits satisfies
\begin{align}\label{eq:int1}
\int_0^1f(x,y)dx &= 2-y+\frac{1}{R_\mathcal{C}}(y-1).
\end{align}
\end{lemma}
\begin{IEEEproof}
Please refer to the proof of \cite[Lemma 7.5]{Measson2006thesis}.
\end{IEEEproof}
For the convolutional code with rate-$1/2$, \eqref{eq:int1} simplifies to
\begin{align}\label{eq:int2}
\int_0^1f(x,y)dx = y.
\end{align}

Now, we are ready to present the first property that gives the relationship between the strength of the component convolutional code and the decoding threshold of the corresponding coupled codes.
\begin{proposition}\label{lem2}
Consider two convolutional codes $\mathcal{C}_1$ and $\mathcal{C}_2$ with their decoders' transfer functions for the information bits, denoted by $f_1(x,y)$ and $f_2(x,y)$, respectively, satisfying
\begin{align}\label{conj_eq1}
\left\{ {\begin{array}{*{20}{c}}
f_1(x,y)<f_2(x,y),\forall x\in(z,1)\\
f_1(x,y)>f_2(x,y),\forall x\in(0,z),
\end{array}} \right.
\end{align}
for some $z \in (0,1)$ and any fixed $y\in (0,1)$. The potential thresholds of the coupled systems based on $\mathcal{C}_1$ and $\mathcal{C}_2$, denoted by $\epsilon_\text{c}(\mathcal{C}_1)$ and $\epsilon_\text{c}(\mathcal{C}_2)$, respectively, satisfy the following condition under the same repetition factor $q<\infty$,
\begin{align}\label{lem2_eq1}
\epsilon_\text{c}(\mathcal{C}_2)>\epsilon_\text{c}(\mathcal{C}_1).
\end{align}
\end{proposition}
\begin{IEEEproof}
See Appendix \ref{app2}.
\end{IEEEproof}

\begin{figure}[t!]
%\begin{minipage}[b]{0.49\linewidth}
	\centering
\includegraphics[width=3.3in,clip,keepaspectratio]{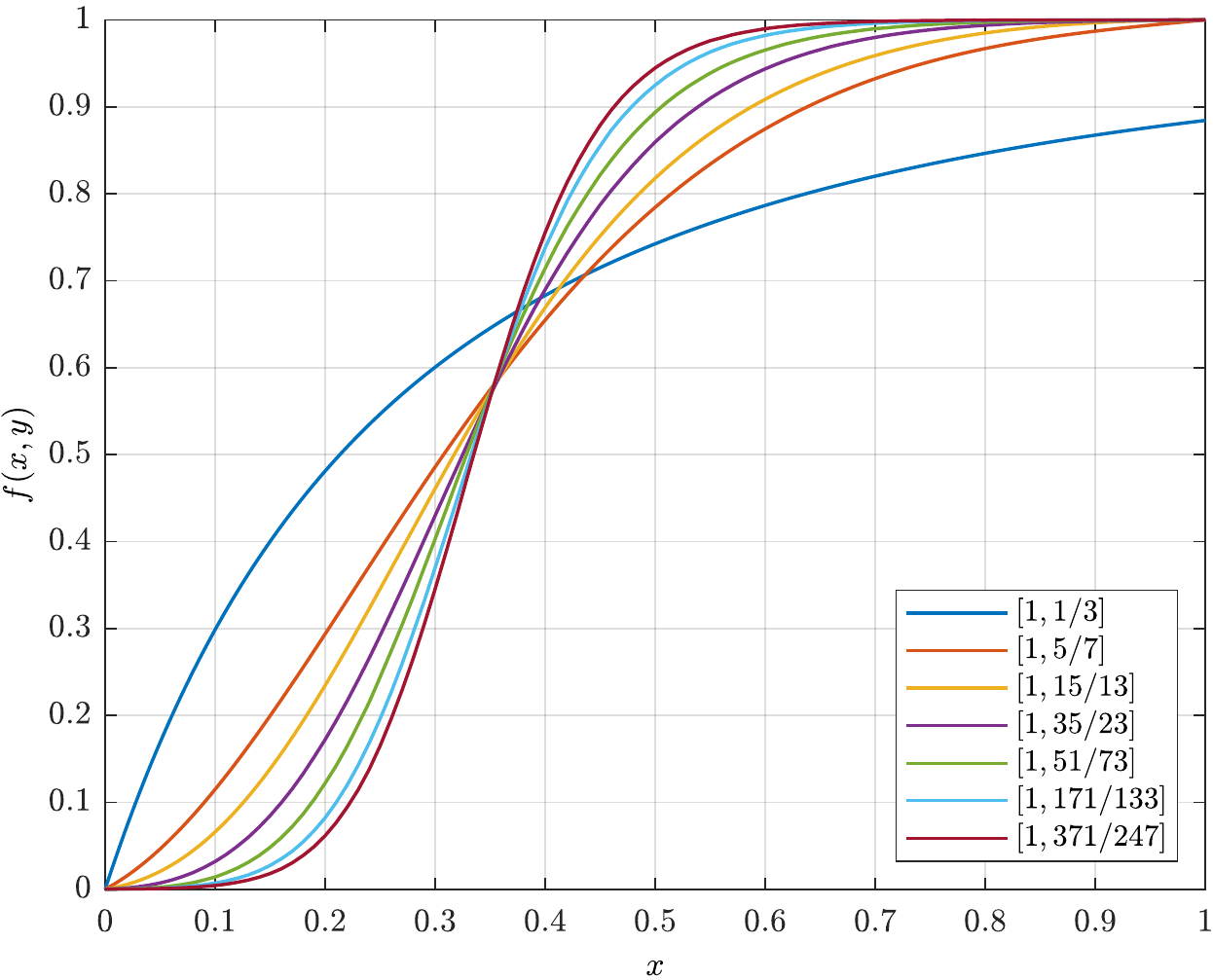}
%\vspace{-6mm}
\caption{Transfer functions of information bits for various convolutional codes.}
\label{fig:trans_diff1}
%\end{minipage}\hfill
\end{figure}

Proposition \ref{lem2} explains the reason why GSC-PCC ensembles built from convolutional codes with a larger number of states have better decoding threshold than those with a lower number of states as reported in Table \ref{MAP1}. This is because a convolutional code with a larger number of states usually achieves a lower bit erasure rate at a lower input erasure probability while achieving a higher bit erasure rate at a higher input erasure probability compared to a convolutional code with a smaller number of states. In Fig. \ref{fig:trans_diff1}, we show the output erasure probability of various transfer functions for $x \in [0,1]$ and $y = 0.66$. One can see that any pair of the considered convolutional codes in the figure satisfying \eqref{conj_eq1}. Thus, when $q$ is fixed and finite, one can use a convolutional code which performs better at a low input erasure probability (not necessarily with a large number of states) to construct a GSC-PCC ensemble with improved decoding threshold. Although we only show one value for $y$ in the figure, we have experimentally verified that the relationships in \eqref{conj_eq1} hold for all the considered convolutional codes with several values of $y \in (0,1)$.

\begin{remark}
If we want to prove that the condition in \eqref{conj_eq1} holds for any pair of convolutional codes, we must explicitly derive and inspect their decoders' transfer functions. However, we can show that $f_1(x,y)$ and $f_2(x,y)$ intersect at $x \in (0,1)$ with a finite number of points. Due to \eqref{eq:int2}, the following holds
\begin{align}
&\int_0^1f_1(x,y)dx = \int_0^1f_2(x,y)dx\\
\Rightarrow &\int_0^1f_1(x,y)-f_2(x,y)dx = 0. \label{eq:contra1}
\end{align}
If $f_1(x,y)$ and $f_2(x,y)$ do not intersect, then it must be true that either $f_1(x,y)>f_2(x,y)$ or $f_1(x,y)<f_2(x,y), \forall x\in (0,1)$. However, this is contradictory to \eqref{eq:contra1}. Hence $f_1(x,y)$ and $f_2(x,y)$ must intersect. In addition, it is impossible for equation $f_1(x,y) = f_2(x,y)$ to have an infinite number of solutions in $x\in (0,1)$ because the transfer function of a convolutional decoder is a rational function whose numerator and denominator are polynomials with finite degrees \cite{1258535}.
\end{remark}

The next property shows the relationship between the decoding threshold of GSC-PCC ensembles, and the repetition factor $q$. Specifically, we investigate the conditions under which the threshold improves with $q$.

\begin{proposition}\label{lem3}
Consider a GSC-PCC ensemble constructed from a convolutional code with decoder transfer function $f(x,y)$. The potential threshold $\epsilon_\text{c}$ improves with $q$ if both of following conditions are satisfied:

1) The fixed point DE equation in \eqref{eq:pot1}, i.e., $f(\epsilon x^{2q-1},1-(1-\epsilon)\rho)=x$, only has two solutions in $x \in (0,1)$ for $\epsilon\in (\epsilon_\text{s},1-R)$, where $\epsilon_\text{s}$ is the BP threshold;

2) The output of the recursive DE equation in \eqref{eq:pot1} with initial condition $x^{(0)}=1$ as $\ell \rightarrow \infty$, i.e., $x^{(\infty)}$, increases with $q$.
\end{proposition}
\begin{IEEEproof}
See Appendix \ref{app3}.
\end{IEEEproof}

\begin{figure}[t!]
%\begin{minipage}[b]{0.49\linewidth}
	\centering
\includegraphics[width=3.3in,clip,keepaspectratio]{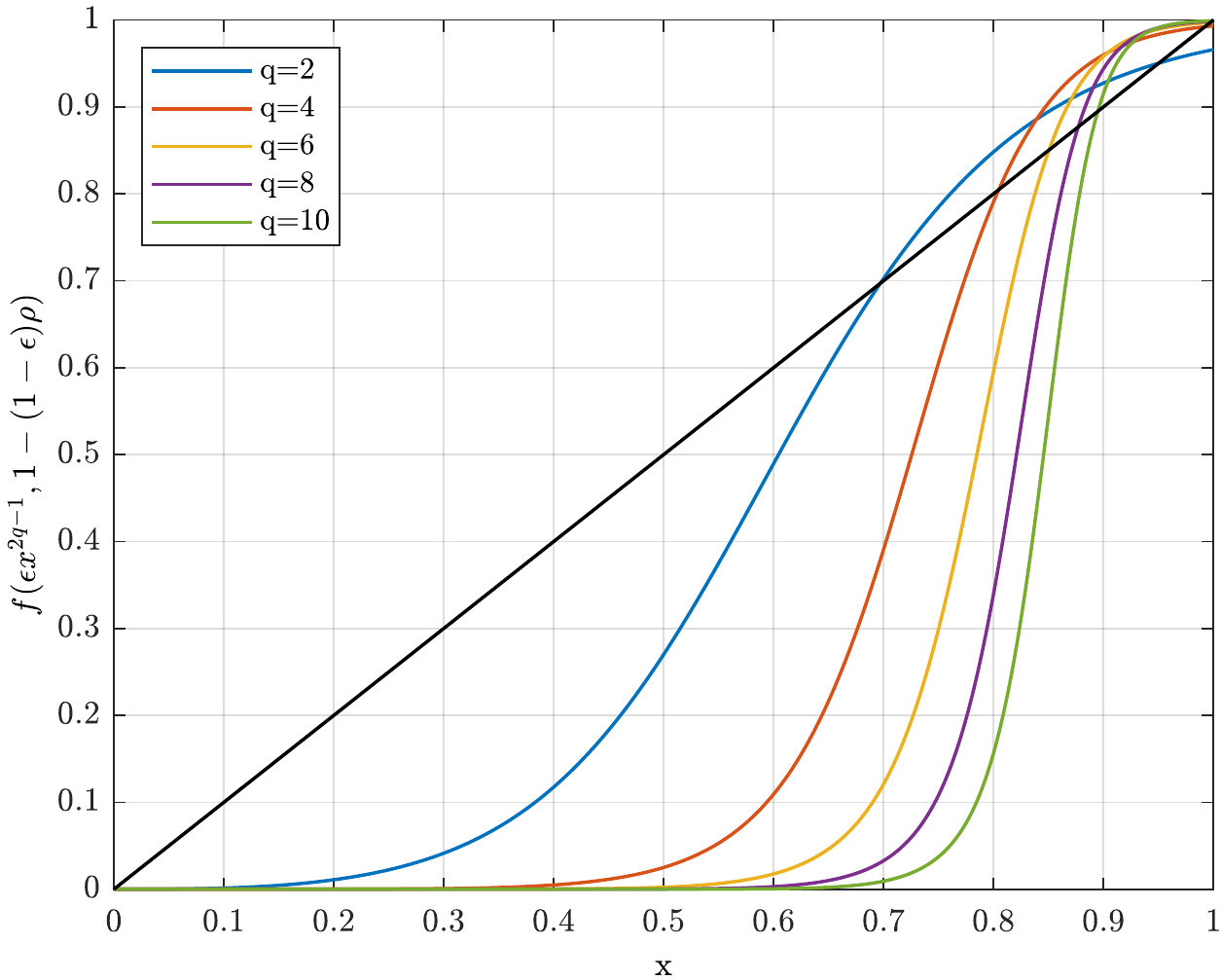}
%\vspace{-6mm}
\caption{Outputs of the transfer functions of an 8-state convolutional code with various $q$ and $R=4/5$.}
\label{fig:trans_func_q}
%\end{minipage}\hfill
%\vspace{-6mm}
\end{figure}

To show that both conditions in Proposition \ref{lem3} hold, we use specific examples. In Fig. \ref{fig:trans_func_q}, we show the values of function $f(\epsilon x^{2q-1},1-(1-\epsilon)\rho)$ for the convolutional code with generator polynomial $(1,15/13)$ for various $q$. In this example, we set $R = 4/5$ and $\epsilon = 0.1698>\epsilon_\text{s}$. One can see that all the curves of the transfer function and line $y=x$ have two intersection points (also known as stationary points according to \cite[Def. 3]{6325197}) while the value of each intersection point increases with $q$. For the ensemble considered in Example \ref{example1}, it can be observed from Fig. \ref{fig:potential_fun_2} that the stationary points of its potential function in \eqref{eq:pot_final1} also increase with $q$. Hence, we expect that the transfer function of any convolutional decoder satisfies both conditions in Proposition \ref{lem3}. To this end, the decoding threshold of general GSC-PCC ensembles can be shown to improve with $q$ until reaching capacity, which is similar to the case considered in Theorem \ref{the:cap}.

\begin{remark}
The potential function in \eqref{eq:pot_final1} is related to that of uncoupled generalized LDPC (GLDPC) codes \cite{1056404}. More precisely, it is associated with the GLDPC codes whose constraint nodes are convolutional codes, e.g., \cite{9174017}. This can be seen by noting that our potential function is a half-iteration shift of the density evolution recursion of an uncoupled GLDPC ensemble by swapping $f$ in \eqref{eq:f} and $g$ in \eqref{eq:g} \cite[Section II-D]{6887298}. Since both coupled systems share many similarities \cite[Lemma 11]{6887298}, the analysis on the potential threshold of our coupled system can be used for the GLDPC counterpart. We also note that the repetition ratio of GSC-PCCs, $\lambda$, can be made irregular, analogous to the irregular variable node degrees of GLDPC codes. However, Tables \ref{table0}-\ref{table1} already show that the BP threshold of GSC-PCCs is close to the corresponding MAP threshold by optimizing $\lambda$ only. Moreover, the analysis in this section demonstrates that regular repetition, i.e., $\lambda = 1/q$, is sufficient to achieve capacity.
\end{remark}

\section{Simulation Results}\label{sec:sim}
In this section, we show the finite length performance of the proposed codes. Unless specified otherwise, we use random interleaving and random parity puncturing (random for each channel realization) in the simulation. In addition, each error point is obtained by collecting at least 300 decoding errors.

\subsection{Performance on the BEC}
\begin{figure}[t!]
%\begin{minipage}[b]{0.49\linewidth}
	\centering
\includegraphics[width=3.3in,clip,keepaspectratio]{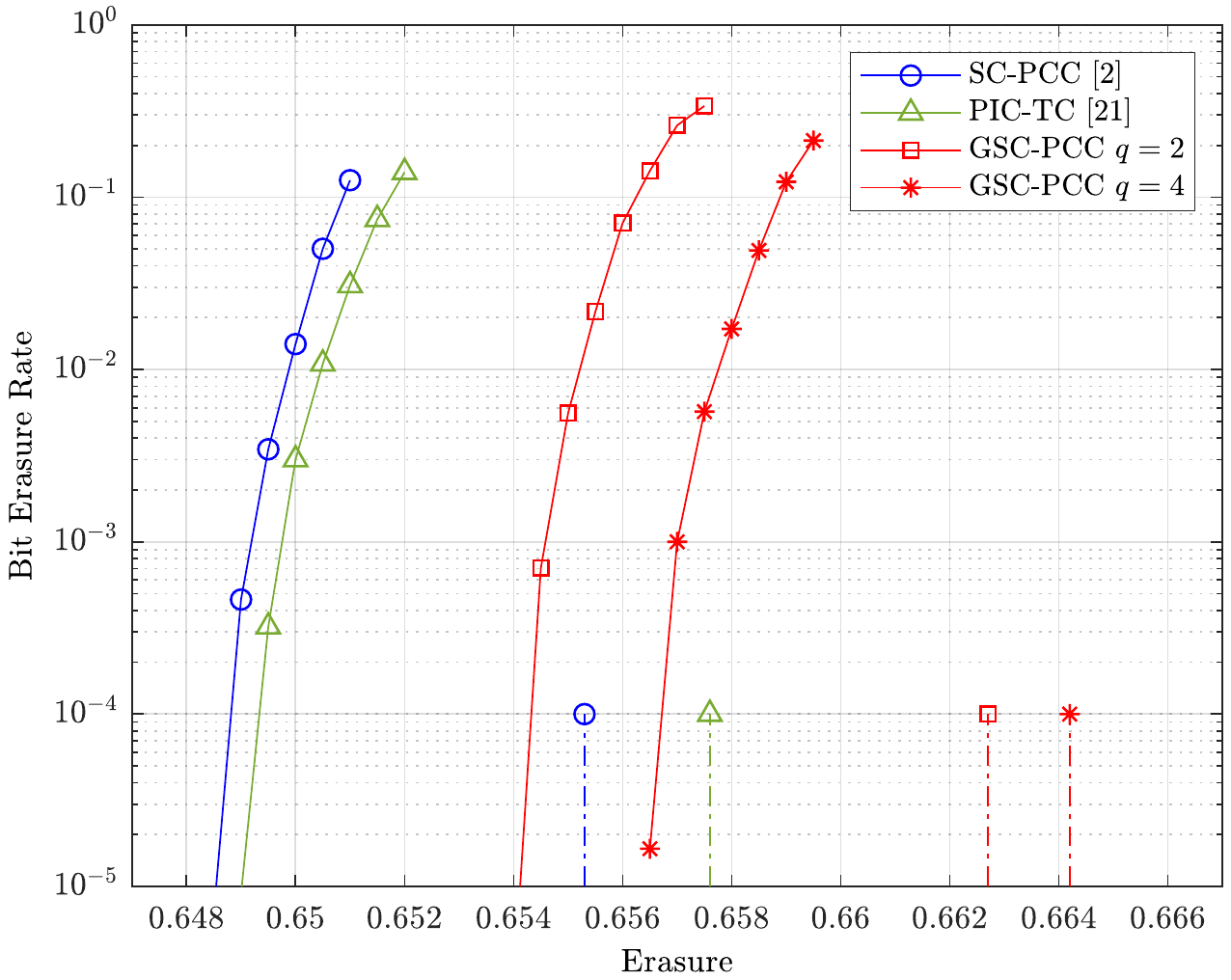}
%\vspace{-6mm}
\caption{BER performance (solid lines) and density evolution thresholds (dash lines) of GSC-PCCs with target rate $1/3$.}
\label{fig:ber}
\end{figure}

We consider GSC-PCCs with identical upper and lower convolutional encoders of generator polynomial $(1,5/7)$. We set $K=10000$, $L=100$, $m=1$, $q\in\{2,4\}$, and $R\in\{1/3,1/2\}$. The values of $\lambda$ are chosen according to Table \ref{table0}. The bit erasure rate (BER) and the BP thresholds for GSC-PCCs are shown in Fig. \ref{fig:ber}. In the same figure, we also plot the BER and decoding thresholds of SC-PCCs \cite{8002601} and PIC-TCs \cite{PIC2020} for comparison purposes. For fair comparison, the benchmark codes and GSC-PCCs have the same target code rate, input message length, coupling length, and coupling memory. To show the best possible performance, all codes are under full decoding of the entire spatial code chain and hence they have the same decoding latency \cite{7296605}.

%\end{minipage}\hfill

\begin{figure}[t!]
%\begin{minipage}[b]{0.49\linewidth}
	\centering
\includegraphics[width=3.3in,clip,keepaspectratio]{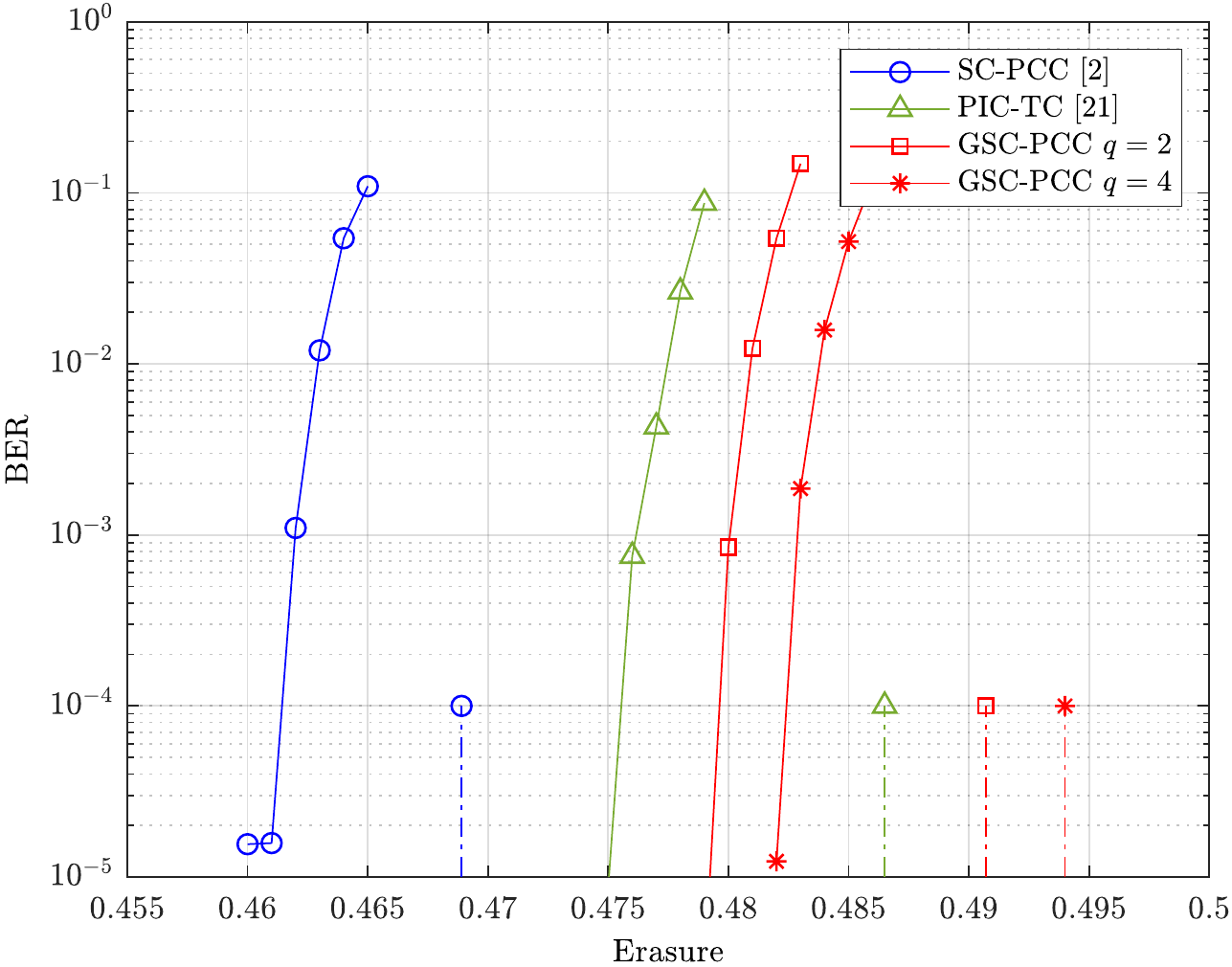}
%\vspace{-6mm}
\caption{BER performance (solid lines) and density evolution thresholds (dash lines) of GSC-PCCs with target rate $1/2$.}
\label{fig:ber2}
%\end{minipage}\hfill
%\vspace{-6mm}
\end{figure}

We observe that for both rates, GSC-PCCs perform better than SC-PCCs and PIC-TCs and the performance gains are in agreement with the DE results. This also confirms that the optimal design of $\lambda$ is effective. It is interesting to see that choosing $q=2$ is sufficient to allow GSC-PCCs outperform SC-PCCs and PIC-TCs while for $q=4$ the proposed codes have a noticeable performance gain over those with $q=2$. Although the BER of uncoupled PCCs is not shown in the figure, one can clearly see that the actual performance of GSC-PCCs at a BER of $10^{-5}$ is much better than the BP thresholds of uncoupled PCCs with the same $q$ (see Table \ref{table1}) or without repetition (see \cite[Table II]{8002601}). It should be noted that the BER performance of GSC-PCCs can be further improved by using a larger $q$ according to our analysis in Section \ref{sec:DE} and Section \ref{sec:TS}.

\subsection{A Criterion For Coupling Bits Selection}\label{sec:coupling_bits_criteria}
From Sections \ref{sec:DE}-\ref{sec:TS}, we know that the excellent threshold is reported for GSC-PCC ensembles which naturally assume random selection of information bits (due to random interleaving). In contrast, the error performance of a GSC-PCC with a fixed code structure can be affected by the selection of coupled information bits.

When the selection of coupling bits is completely random, it is possible that some of the information bits in $\boldsymbol{u}_{t,\text{r}}$ (i.e., the information bits to be repeated) and their $q-1$ replicas can appear in both $\boldsymbol{u}^{\text{U}}_{t,t+j}$ and $\boldsymbol{u}^{\text{L}}_{t,t+j}$ for some $j \in \{0,\ldots,m\}$. In other words, these bits and their $q-1$ replicas are encoded by the upper and lower convolutional component encoders at the same time instant. In this case, these bits cannot benefit from coupling as no extrinsic information from the component codewords at other time instants can be obtained. To enable the exchange of extrinsic information between coupling blocks via these repeated bits, we introduce a simple criterion of selecting coupled bits. That is, each bit in $\boldsymbol{u}_{t,\text{r}}$ and its $q-1$ replicas should not appear in $\boldsymbol{u}^{\text{U}}_{t,t+j}$ and $\boldsymbol{u}^{\text{L}}_{t,t+j}$ at the same time instant, i.e., the repeated bits spread across different time instants. In what follows, we show that by incorporating this criterion in designing GSC-PCCs, a noticeable gain can be attained compared to totally random selection of coupling bits.

\begin{figure}[t!]
%\begin{minipage}[b]{0.49\linewidth}
	\centering
\includegraphics[width=3.3in,clip,keepaspectratio]{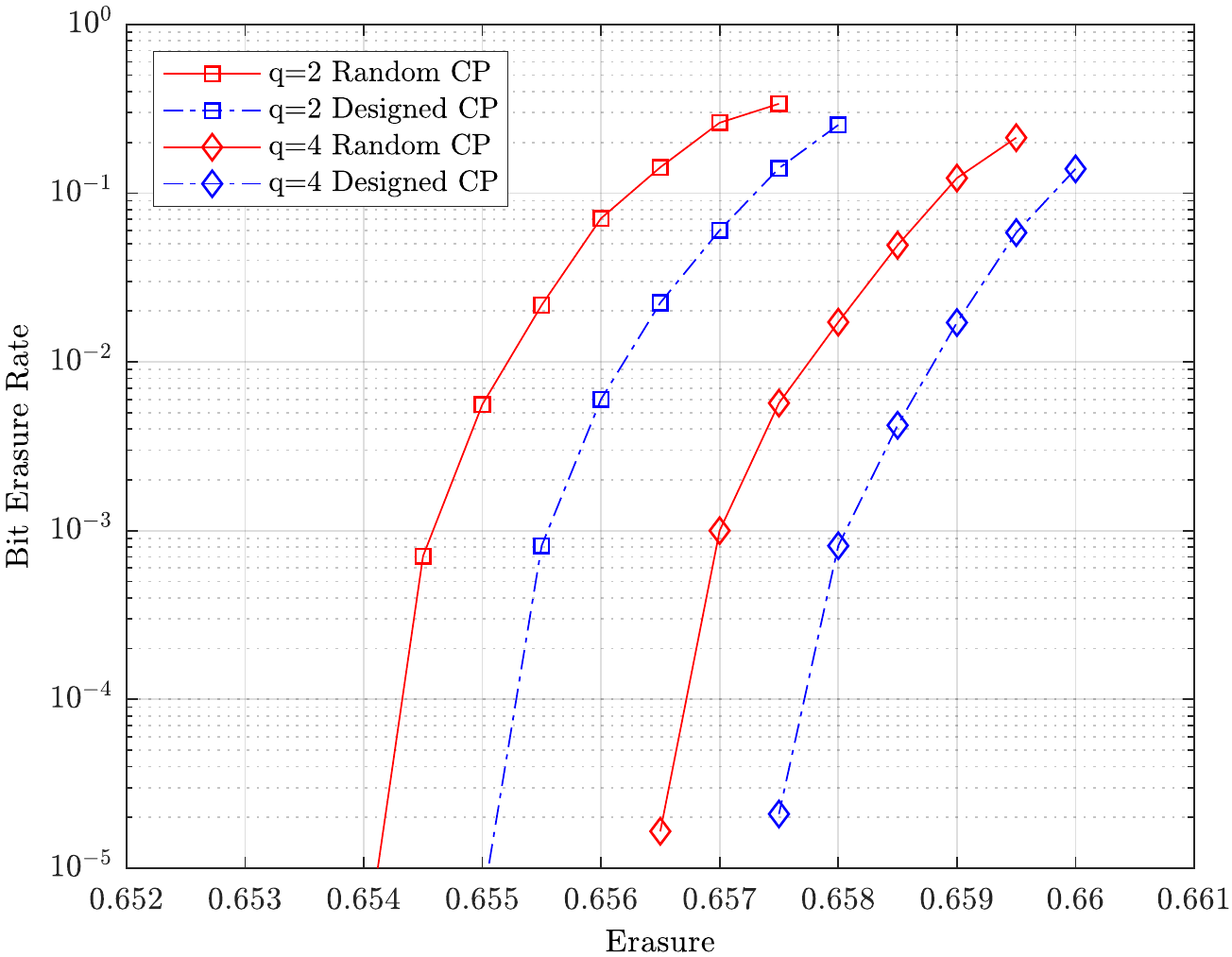}
%\vspace{-6mm}
\caption{BER of GSC-PCCs with target rate $1/3$ and under the proposed criterion.}
\label{fig:BER_designed_CP}
%\end{minipage}\hfill
\end{figure}

We adopt the same settings as in the simulation for Fig. \ref{fig:ber}, except that the employed random interleavers should ensure coupling bits satisfying the aforementioned criterion. The BER performance of the proposed codes under the selected coupling bits (labeled as ``Designed CP'') and that under random selection of coupling bits (labeled as ``Random CP'') is shown in Fig. \ref{fig:BER_designed_CP}. Observe that for both $q=2$ and $q=4$, the error performance is improved.

%0.31 8 state 0.4926

\subsection{Performance on the AWGN Channel}
In this section, we provide the simulation for bit error rate (BER) versus bit signal-to-noise ratio $E_b/N_0$ for GSC-PCCs, PIC-TCs \cite{PIC2020} and SC-LDPC codes \cite{7152893} on the AWGN channel. We have also simulated the frame error rate (FER). Since all FER curves show a similar trend as that for all BER curves, we do not include the FER performance due to the space limitations.

\begin{table}[t!]
  \centering
  %\tiny
 \caption{Five GSC-PCCs Used For Simulations}\label{GSC_setup}
 %\vspace{-3mm}
\begin{tabular}{c c c c c c c}
\hline
 GSC-PCC  & Component Codes &   $\lambda$ & Interleaving  & Puncturing & $\epsilon^{(m=1)}_{\text{BP}}$ \\  \hline
 1 & $(1,5/7)$ & 0.44 & Random & Random  & 0.4907\\
 2 & $(1,15/13)$ & 0.31 & Random & Random & 0.4935 \\
 3 & $(1,15/13)$ & 0.31 & Random & Fixed & 0.4935 \\
 4 & $(1,15/13)$ & 0.31 & Fixed & Fixed  & 0.4935\\
 5 & $(1,15/13)$ & 0.375 & Fixed & Fixed & 0.4928 \\
  \hline
\end{tabular}
%\vspace{-6mm}
\end{table}

\begin{figure}[t!]
%\begin{minipage}[b]{0.49\linewidth}
	\centering
\includegraphics[width=3.3in,clip,keepaspectratio]{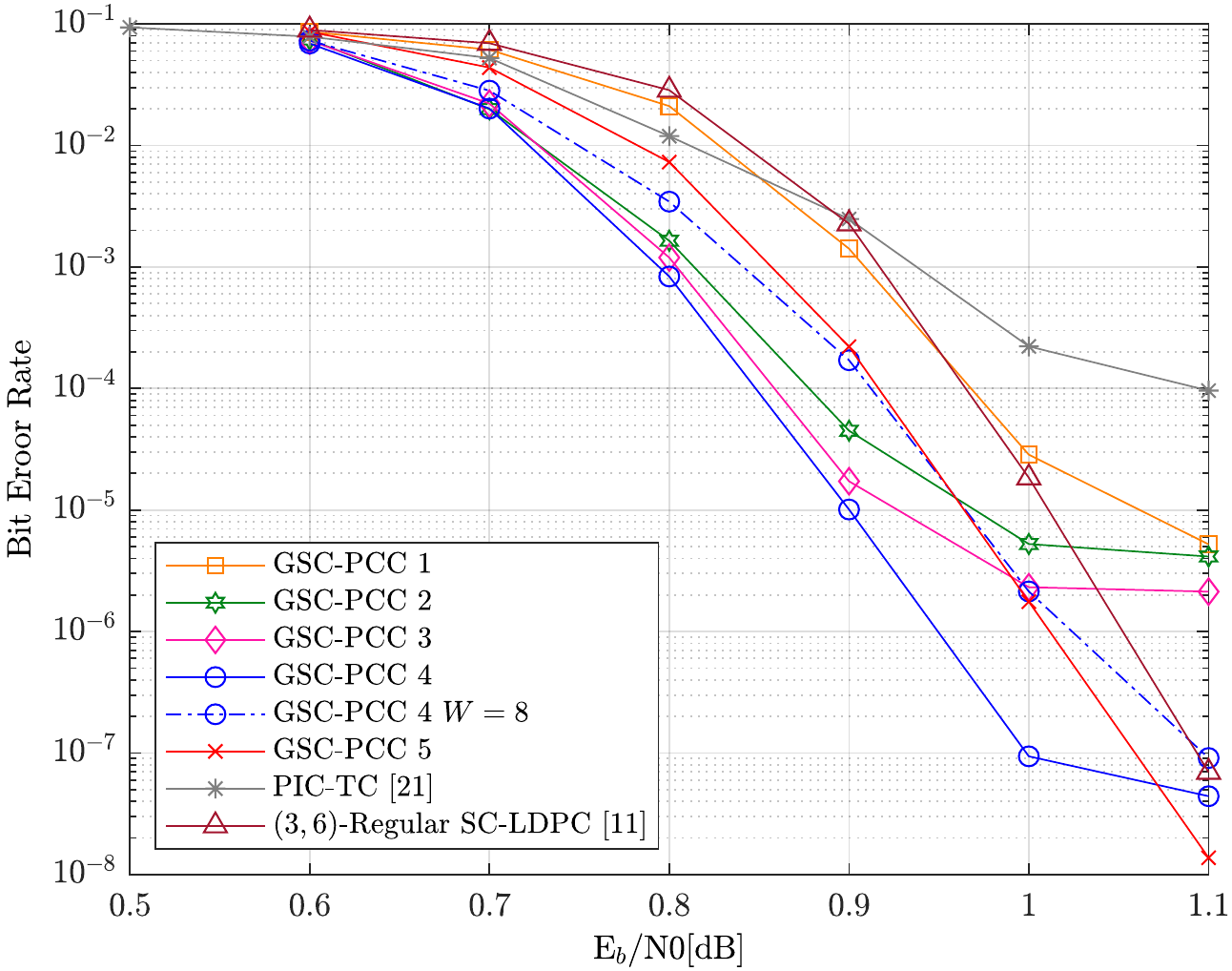}
%\vspace{-6mm}
\caption{BER of GSC-PCCs, PIC-TCs and SC-LDPC with target rate $1/2$.}
\label{fig:ber_res2}
%\end{minipage}\hfill
%\vspace{-6mm}
\end{figure}

First, we consider that all codes have a target rate $R = 1/2$ and coupling length $L=50$. For both GSC-PCCs with $q=2$ and PIC-TCs, we set $K=1000$ and $m=1$. To see the impacts of interleaving, puncturing, and changing of component codes on the finite length performance of GSC-PCCs, we will evaluate the performance of five GSC-PCCs listed in Table \ref{GSC_setup}. Here, for fixed puncturing, we use a periodic puncturing pattern by following \cite[Section VII-A]{7932507}. To obtain the fixed interleavers, we first randomly generate more than 60 sets of interleavers such that the resultant coupling bits satisfy the criterion in Section \ref{sec:coupling_bits_criteria}. Then, we simulate the BER at an $E_b/N_0$ of 1 dB and find the set of interleavers that lead to the lowest BER. The benchmark PIC-TC is with $(1,5/7)$ convolutional component codes, random interleaving and puncturing, and coupling ratio following \cite[Table II]{PIC2020}. The benchmark $(3,6,50,2)$ SC-LDPC code is constructed by following \cite{7152893}, which has a coupling width of 2 and a lifting factor of 1000. The maximum intra-block and inter-block decoding iterations for all turbo-like codes are set to 20 while the maximum BP decoding iterations for SC-LDPC codes are set to 1000. Apart from all the aforementioned codes that are under full decoding, we also showcase an example of the proposed codes, i.e., GSC-PCC 4 in Table \ref{GSC_setup}, using sliding window decoding with a window size $W=8$. The BER versus $E_b/N_0$ is shown in Fig. \ref{fig:ber_res2}.

It can be seen that all the GSC-PCCs outperform the benchmark PIC-TC and SC-LDPC code in terms of waterfall performance on the AWGN channel. Particularly, the GSC-PCC under windowed decoding with a decoding latency of 8000 bits still perform better than the SC-LDPC code and PIC-TC under full decoding with a decoding latency of 50000 bits in the waterfall region. In addition, the GSC-PCCs with a larger BEC decoding threshold reported in Table \ref{GSC_setup} has a better waterfall performance compared to those with a smaller threshold. This means that the excellent performance of the proposed codes on the BEC can be carried over to the AWGN channel. It is also interesting to note that the GSC-PCCs under fixed interleaving and puncturing achieve better waterfall and error floor than their random counterparts. In fact, one can adopt the interleaver designs for turbo-like codes in the literature (e.g., \cite{8214245,9594196}) to attain further performance improvement. Finally, observe that the GSC-PCC with a large $\lambda$ has a lower error floor than that with a small $\lambda$ and comparable error floor performance to the benchmark SC-LDPC code. Therefore, with a small $m$ and finite blocklength, $\lambda$ plays a key role in the trade-off between waterfall and error floor. That said, it is expected that when $m$ becomes large, choosing the maximum $\lambda$, i.e., $\lambda = 1/q$, will result in very good waterfall and error floor.

\section{Conclusions}\label{sec:conclude}
We introduced generalized spatially-coupled parallel concatenated codes, which can be seen as a generalization of the conventional SC-PCCs and have a similar structure to that of PIC-TCs. We derived the density evolution equations for the proposed codes and found the decoding threshold via optimizing the fraction of repeated information bits. By using the potential function argument, we analytically proved that the proposed codes exhibit threshold saturation. Then, we rigorously proved that the GSC-PCC ensemble with 2-state convolutional component codes achieves capacity as the repetition factor tends to infinity and numerically showed that the results can be generalized to GSC-PCC ensembles with other convolutional component codes. To gain more insights into the decoding performance of the proposed codes, the relationships between the strength of the component convolutional codes, decoding threshold of the corresponding GSC-PCCs, and the repetition factor were established. Simulation results of BER under finite blocklength were provided to show that the proposed codes outperform existing class of spatially-coupled codes constructed from component PCCs (or turbo codes).

\appendices
\section{Proof of Theorem \ref{the:cap}}\label{app:cap}
First, the decoder transfer function for the information bits of $(1,1/3)$ convolutional codes can be derived by following \cite{1258535} as
\begin{align}\label{eq:2state}
f_\text{s}(x,y) = \frac{xy(2-2y+xy)}{(1-y+xy)^2}.
\end{align}
Therefore,
\begin{align}
\int_0^{a} f_\text{s}(x z,y)dz = \frac{xy a^2}{xy a-y+1},
\end{align}
and the potential function becomes
\begin{align}
U(x;\epsilon) &= \left(1-\frac{1}{2q}\right) x^{2q}- \frac{\epsilon(1-(1-\epsilon)\rho)x^{4q-2}}{\epsilon(1-(1-\epsilon)\rho)x^{2q-1}-(1-(1-\epsilon)\rho)+1}.
\end{align}
%\vspace{-10mm}
%\par\noindent

Next, we find the necessary condition which $\epsilon \in (0,1)$ has to fulfill such that $U(x;\epsilon)\geq 0, \forall x \in (0,1]$. We have
\begin{subequations}
\begin{align}
U(x;\epsilon)\geq 0\Rightarrow &\;\left(1-\frac{1}{2q}\right) x^{2q}- \frac{\epsilon(1-(1-\epsilon)\rho)x^{4q-2}}{\epsilon(1-(1-\epsilon)\rho)x^{2q-1}-(1-(1-\epsilon)\rho)+1} \geq 0 \\
\Rightarrow &\; \frac{\epsilon(1-(1-\epsilon)\rho)x^{2q-2}}{\epsilon(1-(1-\epsilon)\rho)x^{2q-1}-(1-(1-\epsilon)\rho)+1}\leq \frac{2q-1}{2q} \\
\Rightarrow & \;2q\epsilon(1-\rho+\epsilon\rho)x^{2q-2}-(2q-1)\epsilon(1-\rho+\epsilon\rho)x^{2q-1} \nonumber \\
&+(2q-1)(1-\rho+\epsilon\rho)-(2q-1) \leq 0 \\
\Rightarrow & \;\epsilon(1-\rho+\epsilon\rho)(2qx^{2q-2}-(2q-1)x^{2q-1})-\rho(2q-1)(1-\epsilon) \leq 0 \\
\Rightarrow& \;\epsilon^2x^{2q-2}\left( \frac{2q}{2q-1}-x\right)
+\epsilon\left(x^{2q-2}\left( \frac{2q}{2q-1}-x\right)\frac{1-\rho}{\rho}+1\right)-1 \leq 0 \label{eq:quadratic1}\\
\Rightarrow &\; (\epsilon-\epsilon_1)(\epsilon-\epsilon_2) \leq 0,
\end{align}
\end{subequations}
where $\epsilon_1$ and $\epsilon_2$ are the roots of the quadratic function of \eqref{eq:quadratic1}. Specifically,
\begin{align}
\epsilon_1 &= \frac{-b-\sqrt{b^2-4ac}}{2a} = \frac{-\left(\frac{1-\rho}{\rho}a+1\right)-\sqrt{\left(\frac{1-\rho}{\rho}a+1\right)^2+4a}}{2a}<0, \\
\epsilon_2 &= \frac{-b+\sqrt{b^2-4ac}}{2a} = \frac{-\left(\frac{1-\rho}{\rho}a+1\right)+\sqrt{\left(\frac{1-\rho}{\rho}a+1\right)^2+4a}}{2a} >0 \label{eq:sol2_2s},
\end{align}
where we have used the following definitions for ease of presentation
\begin{align}
&a\triangleq x^{2q-2}\left(\frac{2q}{2q-1}-x\right)>0, \label{notation_a} \\
&b\triangleq x^{2q-2}\left( \frac{2q}{2q-1}-x\right)\frac{1-\rho}{\rho}+1 = \frac{1-\rho}{\rho}a+1>0, \\
&c \triangleq -1.
\end{align}

Since $\epsilon_1<0<\epsilon_2$, it then remains to find $x$ such that $\epsilon_2$ reaches its minimum. This is because we want to ensure that $U(x;\epsilon) \geq 0$ for any $\epsilon<\min_{x\in(0,1]}\epsilon_2$. By taking the following first order partial derivative,
\begin{align}
\frac{\partial\epsilon_2}{\partial x} =& \frac{\partial\epsilon_2}{\partial a}\cdot\frac{\partial a}{\partial x} \nonumber \\
 =& \frac{\sqrt{\left( \frac{1-\rho}{\rho}a+1 \right)^2+4a  }-\frac{1+\rho}{\rho}a-1}{2a^2\sqrt{\left( \frac{1-\rho}{\rho}a+1 \right)^2+4a  }}\cdot\left(\frac{2q(2q-2)}{2q-1}x^{2q-3}-(2q-1)x^{2q-2}\right),
\end{align}
we note that $\epsilon_2$ is strictly decreasing in $x \in \left(0,\frac{4q(q-1)}{2q-1}\right)$ and strictly increasing in $x \in \left(\frac{4q(q-1)}{2q-1},1\right]$. This can be seen by first noting that the partial derivative $\frac{\partial\epsilon_2}{\partial a}$ satisfies
\begin{align}
\frac{\partial\epsilon_2}{\partial a} = \sqrt{\left( \frac{1-\rho}{\rho}a+1 \right)^2+4a  }-\frac{1+\rho}{\rho}a-1<0,
\end{align}
due to the fact that
\begin{align}
 \left( \frac{1-\rho}{\rho}a+1 \right)^2+4a-\left(\frac{1+\rho}{\rho}a+1\right)^2 = -\frac{4a^2}{\rho} <0.
\end{align}
In addition, it is easy to see that the partial derivative $\frac{\partial a}{\partial x} = \frac{2q(2q-2)}{2q-1}x^{2q-3}-(2q-1)x^{2q-2}$ is strictly increasing in $x \in \left(0,\frac{4q(q-1)}{2q-1}\right)$ and strictly decreasing in $x \in \left(\frac{4q(q-1)}{2q-1},1\right]$. Therefore,
\begin{align}\label{eq:s2_proof_2}
x=\argmin_{x\in(0,1]} \epsilon_2 = \frac{4q(q-1)}{(2q-1)^2}.
\end{align}
Note that in order to ensure $x>0$, one should have $q\geq2$ ($q=1$ corresponds to the case of SC-PCCs \cite{8002601}).

The potential threshold can be obtained by substituting \eqref{eq:s2_proof_2} into \eqref{eq:sol2_2s},
\begin{subequations}
\begin{align}
\epsilon_{\text{c}} =\epsilon_2=&\frac{-\left(\frac{1-\rho}{\rho}a+1\right)+\sqrt{\left(\frac{1-\rho}{\rho}a+1\right)^2+4a}}{2a}  \\
=& -\left(\frac{1-\rho}{2\rho}+\frac{1}{2a}\right)+\sqrt{\frac{(a+\rho)^2+2a\rho(\rho-a)+a^2\rho^2}{4a^2\rho^2}}  \\
\geq& -\left(\frac{1-\rho}{2\rho}+\frac{1}{2a}\right)+\sqrt{\frac{(a+\rho)^2+2a\rho(\rho-a)+a^2\rho^2\left(\frac{\rho-a}{\rho+a}\right)^2}{4a^2\rho^2}} \\
=& -\left(\frac{1-\rho}{2\rho}+\frac{1}{2a}\right)+\sqrt{\frac{\left((a+\rho)+\frac{\rho-a}{\rho+a}a\rho\right)^2}{4a^2\rho^2}} \\
=& -\frac{a+\rho-a\rho}{2a\rho}+\frac{(a+\rho)+\frac{\rho-a}{\rho+a}a\rho}{2a\rho} \\
=& \frac{\rho}{a+\rho} \\
%=& \frac{\frac{R_0(1-R)}{qR(1-R_0)}}{\frac{2q}{(2q-1)^2}\left(\frac{4q(q-1)}{(2q-1)^2}\right)^{2q-2} +\frac{R_0(1-R)}{qR(1-R_0)}}\nonumber \\
%=& \frac{\frac{1-R}{2R}}{\frac{2q^2}{(2q-1)^2}\left(\frac{4q(q-1)}{(2q-1)^2}\right)^{2q-2} +\frac{1-R}{2R}}\nonumber \\
=& (1-R)\left(1-\frac{1}{1+\frac{1}{R(2qa-1)}}\right)\label{eq:rho_use}\\
\geq& (1-R)\left(1-\frac{R}{R+q}\right),\label{eq:state2_final_step}
\end{align}
\end{subequations}
where in \eqref{eq:rho_use} we have used $\rho = \frac{R_0(1-R)}{qR(1-R_0)}=\frac{1-R}{2qR}$ and \eqref{eq:state2_final_step} follows from
\begin{subequations}
\begin{align}
\frac{(q+1)(q-\frac{1}{2})^{4q-2}}{q^{2q+1}(q-1)^{2q-2}} =& \frac{(q+1)(q-1)}{q^2}\left(\frac{(q-\frac{1}{2})^2}{q(q-1)}\right)^{2q-1}  \\
=& \left(1-\frac{1}{q^2} \right)\left(1+\frac{1}{4(q^2-q)}\right)^{2q-1} \label{ineq1_before} \\
\geq& 1\label{ineq1},
\end{align}
\end{subequations}
which implies that
\begin{subequations}
\begin{align}
&\frac{2^{4q-2}q^{2q}(q-1)^{2q-2}}{(2q-1)^{4q-2}} \leq \frac{q+1}{q}  \\
\Rightarrow & 2qa-1\leq \frac{1}{q} \qquad (\text{by} \; \eqref{notation_a} \;\& \;\eqref{eq:s2_proof_2}),
\end{align}
\end{subequations}
and \eqref{ineq1} holds because by inspecting the derivative of \eqref{ineq1_before}, i.e.,
\begin{align}
\frac{4\left(1+\frac{1}{4(q^2-q)}\right)^{2q}(q-1)^2\left( (2q+2q^2)\ln\left(1+\frac{1}{4(q^2-q)}\right)-1\right)}{\left((2q-1)q\right)^2},
\end{align}
we note that the function \eqref{ineq1_before} is strictly increasing in $q \in [2,2.91486)$ and strictly decreasing in $q \in (2.91486,\infty)$ such that its minimum is achieved when $q \rightarrow \infty$.

Using Theorem \ref{the:ts}, we conclude that the BP threshold of the considered GSC-PCC ensembles with $L \rightarrow \infty$, $m \rightarrow \infty$ and $L \gg m$ is lower bounded by \eqref{eq:state2_final_step}. This completes the proof.

\section{Proof of Proposition \ref{lem2}}\label{app2}
We first show that the following inequality is true.
\begin{align}\label{the_f}
\int_{0}^{\vartheta}f_1(x,y)dx > \int_{0}^{\vartheta}f_2(x,y)dx, \forall \vartheta \in (0,1).
\end{align}
It is immediate that \eqref{the_f} holds for $\vartheta \in (0,z]$ due to the second equality of \eqref{conj_eq1}. As for $\vartheta \in (z,1)$, we have
\begin{align}
\int_{0}^{\vartheta}f_1(x,y)dx =& \int_{0}^{1}f_1(x,y)dx -  \int_{\vartheta}^{1}f_1(x,y)dx \nonumber \\
\overset{\eqref{eq:int2}}{=}& \int_{0}^{1}f_2(x,y)dx -  \int_{\vartheta}^{1}f_1(x,y)dx \nonumber \\
 \overset{\eqref{conj_eq1}}{>}& \int_{0}^{1}f_2(x,y)dx-  \int_{\vartheta}^{1}f_2(x,y)dx \nonumber \\
 =& \int_{0}^{\vartheta}f_2(x,y)dx.
\end{align}
For the transfer functions satisfying \eqref{the_f}, we can show that the potential functions in relation to $\mathcal{C}_1$ and $\mathcal{C}_2$ satisfy the following condition for any $x \in (0,1]$ and $\epsilon \in (0,1)$.
\begin{align}\label{ineq:pot}
U(x,\epsilon)(\mathcal{C}_2) =& \left(1-\frac{1}{2q}\right) x^{2q}- \int_0^{x^{2q-1}} f_2(\epsilon z,1-(1-\epsilon)\rho)dz \nonumber \\
=& \left(1-\frac{1}{2q}\right) x^{2q}- \frac{1}{\epsilon}\int_0^{\epsilon x^{2q-1}} f_2(z',1-(1-\epsilon)\rho)dz'\qquad \left(z' = \epsilon z\right) \nonumber \\
\overset{\eqref{the_f}}{>}& \left(1-\frac{1}{2q}\right) x^{2q}- \frac{1}{\epsilon}\int_0^{\epsilon x^{2q-1}} f_1(z',1-(1-\epsilon)\rho)dz'   \nonumber \\
=& U(x,\epsilon)(\mathcal{C}_1).
\end{align}
The inequality in \eqref{ineq:pot} implies that $\min_{x \in (0,1]} U(x,\epsilon)(\mathcal{C}_1)= \min_{x \in (0,1]}U(x,\epsilon')(\mathcal{C}_2)=0,\exists\epsilon'>\epsilon$. As a result, we obtain \eqref{lem2_eq1} by using Definition \ref{def:2}.

\section{Proof of Proposition \ref{lem3}}\label{app3}
Consider that the two solutions, $x_1$ and $x_2$, satisfy $0<x_1<x_2<1$. We first show that the following holds.
\begin{align}\label{conj_eq2}
\left\{ {\begin{array}{*{20}{c}}
&f(\epsilon x^{2q-1},1-(1-\epsilon)\rho)<x,\forall x\in(0,x_1)\cup(x_2,1]\\
&f(\epsilon x^{2q-1},1-(1-\epsilon)\rho)>x,\forall x\in(x_1,x_2)
\end{array}} \right..
\end{align}

Recall that the transfer function of any convolutional decoder is strictly increasing in all its arguments \cite[Lemma 1]{8002601}. Thus, it is easy to see that $f(\epsilon x^{2q-1},1-(1-\epsilon)\rho)$ is also strictly increasing in $x\in (0,1]$. Since $x > f(\epsilon x^{2q-1},1-(1-\epsilon)\rho)$ for $x=1$ and by realizing that $x_2$ is the largest root of $x = f(\epsilon x^{2q-1},1-(1-\epsilon)\rho)$, we have
\begin{align}\label{eq:region1}
x > f(\epsilon x^{2q-1},1-(1-\epsilon)\rho), \forall x \in (x_2,1].
\end{align}

Since $x_1$ is the smallest non-zero root, then for $\forall x \in(0,x_1)$, one must have either $x<f(\epsilon x^{2q-1},1-(1-\epsilon)\rho)$ or $x>f(\epsilon x^{2q-1},1-(1-\epsilon)\rho)$. If the former holds, then the following must be true
\begin{align}\label{eq:former_contra}
&x^{(\ell)} = f(\epsilon (x^{(\ell-1)})^{2q-1},1-(1-\epsilon)\rho)>x^{(\ell-1)},  \\
\Rightarrow & x^{(\infty)} =f(\epsilon (x^{(\infty)})^{2q-1},1-(1-\epsilon)\rho) = x_1>0.
\end{align}
This means that even given an initial condition very close to 0, i.e., $x^{(0)}\rightarrow 0$, the iterative system defined by the recursion in \eqref{eq:former_contra} will never converge to 0 as $\ell \rightarrow \infty$, which is not true. Hence, one can only have the following
\begin{align}\label{eq:region2}
x > f(\epsilon x^{2q-1},1-(1-\epsilon)\rho), \forall x \in (0,x_1).
\end{align}

However, there must exist a region on which $x < f(\epsilon x^{2q-1},1-(1-\epsilon)\rho)$ because the condition $\epsilon > \epsilon_\text{s}$ leads to $U'(x;\epsilon)\leq0, \exists x\in (0,1)$ according to Definition \ref{def:sst} and $f(\epsilon x^{2q-1},1-(1-\epsilon)\rho)$ is increasing with $\epsilon$. The only possible region for such condition to hold is $x \in (x_1,x_2)$. This leads to \eqref{conj_eq2}.

As for the largest root, the following condition holds due to Definition \ref{def:2} and \eqref{eq:region1}
\begin{align}\label{eq:con_x2}
%&U'(x_2;\epsilon) = 0, \nonumber \\
 x_2=\argmin_{x \in (0,1]} U(x;\epsilon)  %\nonumber \\
=x^{(\infty)}=f(\epsilon (x^{(\infty)})^{2q-1},1-(1-\epsilon)\rho), \;x^{(0)}=1.
\end{align}
Using \eqref{conj_eq2} and \eqref{eq:con_x2}, we obtain the following system of equations by letting $\epsilon = \epsilon_{\text{c}}$%letting $x = x_2$, the following holds as $\epsilon = \epsilon_{\text{c}}$ according to \cite[Def. 3]{6325197},
\begin{align}\label{conj_eq3}
\left\{ {\begin{array}{*{20}{c}}
U(x_2,\epsilon_{\text{c}}) = 0\\
U'(x_2,\epsilon_{\text{c}}) = 0
\end{array}} \right. \Rightarrow \left\{ {\begin{array}{*{20}{c}}
\left(1-\frac{1}{2q}\right)x_2^{2q}-\int_0^{x_2^{2q-1}}f(\epsilon_{\text{c}} z,1-(1-\epsilon_{\text{c}})\rho)dz = 0\\
x_2 = f(\epsilon_{\text{c}} x_2^{2q-1},1-(1-\epsilon_{\text{c}})\rho) = 0
\end{array}} \right..
\end{align}
Given a specific transfer function $f$, one can solve for $\epsilon_{\text{c}}$ as a function of $q$ from the above equations. Since the transfer function of any convolutional decoder cannot be expressed as a universal closed form, we instead look at the following derivative
\begin{align}
\frac{\partial \epsilon_{\text{c}}(q)}{\partial q} = \frac{\partial \epsilon_{\text{c}}(x_2,q)}{\partial x_2}\cdot \frac{\partial x_2(\epsilon_{\text{c}},q)}{\partial q},
\end{align}
where $\epsilon_{\text{c}}(x_2,q)$ is the solution of the first equation in \eqref{conj_eq3}, and $ x_2(\epsilon_{\text{c}},q)$ is the solution of the second equation in \eqref{conj_eq3}. Consider $x'_2 \in (x_2,1)$. Then, the following holds
\begin{subequations}
\begin{align}
\eqref{eq:region1}\Rightarrow&U'(x;\epsilon)>0,\forall x\in(x_2,1] \\
\Rightarrow & U(x'_2;\epsilon_{\text{c}}(x_2,q))>U(x_2;\epsilon_{\text{c}}(x_2,q)) = U(x'_2;\epsilon_{\text{c}}(x'_2,q))=0  \\
\Rightarrow & \epsilon_{\text{c}}(x'_2,q)>\epsilon_{\text{c}}(x_2,q)\label{eq:conj4} \\
\Rightarrow & \frac{\partial \epsilon_{\text{c}}(x_2,q)}{\partial x_2}>0,
%\Rightarrow &\left(1-\frac{1}{2q}\right)(x'_2)^{2q}-\int_0^{(x'_2)^{2q-1}}f(\epsilon_{\text{c}} z,1-(1-\epsilon_{\text{c}})\rho)dz>0. \label{eq:conj4}
\end{align}
\end{subequations}
where \eqref{eq:conj4} follows from the fact that $U(x;\epsilon)$ is strictly decreasing in $\epsilon \in (0,1]$ \cite{6325197,6887298}. In addition, with \eqref{eq:con_x2} and condition 2), i.e., $x^{(\infty)}$, increases with $q$, it is immediate that $\frac{\partial x_2(\epsilon_{\text{c}},q)}{\partial q}>0$. Therefore,
$\frac{\partial \epsilon_{\text{c}}(q)}{\partial q}>0$, which means that the potential threshold $\epsilon_\text{c}$ improves with $q$.

\bibliographystyle{IEEEtran}
\bibliography{MinQiu}

\begin{thebibliography}{10}
\providecommand{\url}[1]{#1}
\csname url@samestyle\endcsname
\providecommand{\newblock}{\relax}
\providecommand{\bibinfo}[2]{#2}
\providecommand{\BIBentrySTDinterwordspacing}{\spaceskip=0pt\relax}
\providecommand{\BIBentryALTinterwordstretchfactor}{4}
\providecommand{\BIBentryALTinterwordspacing}{\spaceskip=\fontdimen2\font plus
\BIBentryALTinterwordstretchfactor\fontdimen3\font minus
  \fontdimen4\font\relax}
\providecommand{\BIBforeignlanguage}[2]{{%
\expandafter\ifx\csname l@#1\endcsname\relax
\typeout{** WARNING: IEEEtran.bst: No hyphenation pattern has been}%
\typeout{** loaded for the language `#1'. Using the pattern for}%
\typeout{** the default language instead.}%
\else
\language=\csname l@#1\endcsname
\fi
#2}}
\providecommand{\BIBdecl}{\relax}
\BIBdecl

\bibitem{GSCPCC2021ISIT}
M.~Qiu, X.~Wu, J.~Yuan, and A.~Graell~i Amat, ``Generalized spatially coupled
  parallel concatenated convolutional codes with partial repetition,'' in
  \emph{Proc. IEEE Int. Symp. Inf. Theory (ISIT)}, Jul. 2021, pp. 581--586.

\bibitem{8002601}
S.~Moloudi, M.~Lentmaier, and A.~{Graell i Amat}, ``Spatially coupled
  turbo-like codes,'' \emph{IEEE Trans. Inf. Theory}, vol.~63, no.~10, pp.
  6199--6215, Oct. 2017.

\bibitem{397441}
C.~{Berrou}, A.~{Glavieux}, and P.~{Thitimajshima}, ``Near shannon limit
  error-correcting coding and decoding: {T}urbo-codes,'' in \emph{Proc. IEEE
  Int. Conf. Commun. (ICC)}, vol.~2, May 1993, pp. 1064--1070.

\bibitem{Vucetic:2000:TCP:352869}
B.~Vucetic and J.~Yuan, \emph{Turbo Codes: Principles and Applications}.\hskip
  1em plus 0.5em minus 0.4em\relax Norwell, MA, USA: Kluwer Academic
  Publishers, 2000.

\bibitem{Gallager63low-densityparity-check}
R.~G. Gallager, ``Low-density parity-check codes,'' \emph{MIT Press}, 1963.

\bibitem{782171}
A.~{Jimenez Felstrom} and K.~S. {Zigangirov}, ``Time-varying periodic
  convolutional codes with low-density parity-check matrix,'' \emph{IEEE Trans.
  Inf. Theory}, vol.~45, no.~6, pp. 2181--2191, Sep. 1999.

\bibitem{1056404}
R.~Tanner, ``A recursive approach to low complexity codes,'' \emph{IEEE Trans.
  Inf. Theory}, vol.~27, no.~5, pp. 533--547, Sep. 1981.

\bibitem{5571910}
M.~Lentmaier, A.~Sridharan, D.~J. Costello, and K.~S. Zigangirov, ``Iterative
  decoding threshold analysis for {LDPC} convolutional codes,'' \emph{IEEE
  Trans. Inf. Theory}, vol.~56, no.~10, pp. 5274--5289, Oct. 2010.

\bibitem{5695130}
S.~Kudekar, T.~J. Richardson, and R.~L. Urbanke, ``Threshold saturation via
  spatial coupling: Why convolutional {LDPC} ensembles perform so well over the
  {BEC},'' \emph{IEEE Trans. Inf. Theory}, vol.~57, no.~2, pp. 803--834, Feb.
  2011.

\bibitem{6589171}
S.~{Kudekar}, T.~{Richardson}, and R.~L. {Urbanke}, ``Spatially coupled
  ensembles universally achieve capacity under belief propagation,'' \emph{IEEE
  Trans. Inf. Theory}, vol.~59, no.~12, pp. 7761--7813, Dec. 2013.

\bibitem{7152893}
D.~G.~M. {Mitchell}, M.~{Lentmaier}, and D.~J. {Costello}, ``Spatially coupled
  {LDPC} codes constructed from protographs,'' \emph{IEEE Trans. Inf. Theory},
  vol.~61, no.~9, pp. 4866--4889, Sep. 2015.

\bibitem{Smith12}
B.~P. Smith, A.~Farhood, A.~Hunt, F.~R. Kschischang, and J.~Lodge, ``Staircase
  codes: {FEC} for 100 {G}b/s {OTN},'' \emph{J. Lightw. Technol.}, vol.~30,
  no.~1, pp. 110--117, Jan. 2012.

\bibitem{910572}
F.~R. Kschischang, B.~J. Frey, and H.~A. Loeliger, ``Factor graphs and the
  sum-product algorithm,'' \emph{IEEE Trans. Inf. Theory}, vol.~47, no.~2, pp.
  498--519, Feb. 2001.

\bibitem{669119}
S.~{Benedetto}, D.~{Divsalar}, G.~{Montorsi}, and F.~{Pollara}, ``Serial
  concatenation of interleaved codes: performance analysis, design, and
  iterative decoding,'' \emph{IEEE Trans. Inf. Theory}, vol.~44, no.~3, pp.
  909--926, May 1998.

\bibitem{5361461}
W.~{Zhang}, M.~{Lentmaier}, K.~S. {Zigangirov}, and D.~J. {Costello}, ``Braided
  convolutional codes: A new class of turbo-like codes,'' \emph{IEEE Trans.
  Inf. Theory}, vol.~56, no.~1, pp. 316--331, Jan. 2010.

\bibitem{8631116}
S.~{Moloudi}, M.~{Lentmaier}, and A.~{Graell i Amat}, ``Spatially coupled
  turbo-like codes: A new trade-off between waterfall and error floor,''
  \emph{IEEE Trans. Commun.}, vol.~67, no.~5, pp. 3114--3123, 2019.

\bibitem{9448689}
M.~Mahdavi, M.~Umar~Farooq, L.~Liu, O.~Edfors, V.~Öwall, and M.~Lentmaier,
  ``The effect of coupling memory and block length on spatially coupled
  serially concatenated codes,'' in \emph{Proc. IEEE VTC-Spring}, Apr. 2021,
  pp. 1--7.

\bibitem{8368318}
L.~Yang, Y.~Xie, X.~Wu, J.~Yuan, X.~Cheng, and L.~Wan, ``Partially
  information-coupled turbo codes for {LTE} systems,'' \emph{IEEE Trans.
  Commun.}, vol.~66, no.~10, pp. 4381--4392, Oct. 2018.

\bibitem{4907407}
A.~Larmo, M.~Lindström, M.~Meyer, G.~Pelletier, J.~Torsner, and H.~Wiemann,
  ``The {LTE} link-layer design,'' \emph{IEEE commun. Mag.}, vol.~47, no.~4,
  pp. 52--59, Apr. 2009.

\bibitem{8989359}
M.~Qiu, X.~Wu, and J.~Yuan, ``Density evolution analysis of partially
  information coupled turbo codes on the erasure channel,'' in \emph{Inf.
  Theory Workshop (ITW)}, Aug. 2019, pp. 1--5.

\bibitem{PIC2020}
M.~Qiu, X.~Wu, A.~{Graell i Amat}, and J.~Yuan, ``Analysis and design of
  partially information- and partially parity-coupled turbo codes,'' \emph{IEEE
  Trans. Commun.}, vol.~69, no.~4, pp. 2107--2122, Apr. 2021.

\bibitem{8301547}
L.~{Yang}, Y.~{Xie}, J.~{Yuan}, X.~{Cheng}, and L.~{Wan}, ``Chained {LDPC}
  codes for future communication systems,'' \emph{IEEE Commun. Lett.}, vol.~22,
  no.~5, pp. 898--901, 2018.

\bibitem{9491085}
X.~Wu, M.~Qiu, and J.~Yuan, ``Partially information coupled bit-interleaved
  polar coded modulation,'' \emph{IEEE Trans. Commun.}, vol.~69, no.~10, pp.
  6409--6423, Oct. 2021.

\bibitem{9174156}
X.~{Wu}, M.~{Qiu}, and J.~{Yuan}, ``Partially information coupled duo-binary
  turbo codes,'' in \emph{Proc. IEEE Int. Symp. Inf. Theory (ISIT)}, 2020, pp.
  461--466.

\bibitem{Measson2006thesis}
C.~Measson, ``Conservation laws for coding,'' Ph.D. dissertation, {\'E}cole
  polytechnique f{\'e}d{\'e}rale de Lausanne, Lausanne, Switzerland, 2006.

\bibitem{6325197}
A.~{Yedla}, Y.~{Jian}, P.~S. {Nguyen}, and H.~D. {Pfister}, ``A simple proof of
  threshold saturation for coupled scalar recursions,'' in \emph{Proc. Int.
  Symp. Turbo Codes Iterative Inf. Process (ISTC)}, 2012, pp. 51--55.

\bibitem{5308225}
N.~{Pillay}, H.~{Xu}, and F.~{Takawira}, ``Dual-repeat-punctured turbo codes on
  {AWGN} channels,'' in \emph{Proc. IEEE AFRICON}, 2009, pp. 1--6.

\bibitem{1055186}
L.~Bahl, J.~Cocke, F.~Jelinek, and J.~Raviv, ``Optimal decoding of linear codes
  for minimizing symbol error rate,'' \emph{IEEE Trans. Inf. Theory}, vol.~20,
  no.~2, pp. 284--287, Mar. 1974.

\bibitem{Richardson:2008:MCT:1795974}
T.~Richardson and R.~Urbanke, \emph{Modern Coding Theory}.\hskip 1em plus 0.5em
  minus 0.4em\relax New York, NY, USA: Cambridge Univ. Press, 2008.

\bibitem{370145}
M.~R. {Best}, M.~V. {Burnashev}, Y.~{Levy}, A.~{Rabinovich}, P.~C. {Fishburn},
  A.~R. {Calderbank}, and D.~J. {Costello}, ``On a technique to calculate the
  exact performance of a convolutional code,'' \emph{IEEE Trans. Inf. Theory},
  vol.~41, no.~2, pp. 441--447, 1995.

\bibitem{1258535}
B.~M. {Kurkoski}, P.~H. {Siegel}, and J.~K. {Wolf}, ``Exact probability of
  erasure and a decoding algorithm for convolutional codes on the binary
  erasure channel,'' in \emph{Proc. IEEE Globecom}, vol.~3, Dec. 2003, pp.
  1741--1745.

\bibitem{7353121}
D.~G.~M. {Mitchell}, M.~{Lentmaier}, A.~E. {Pusane}, and D.~J. {Costello},
  ``Randomly punctured {LDPC} codes,'' \emph{IEEE J. Sel. Areas Commun.},
  vol.~34, no.~2, pp. 408--421, 2016.

\bibitem{1523540}
C.~{Measson}, R.~{Urbanke}, A.~{Montanari}, and T.~{Richardson}, ``Maximum a
  posteriori decoding and turbo codes for general memoryless channels,'' in
  \emph{Proc. IEEE Int. Symp. Inf. Theory (ISIT)}, Sep. 2005, pp. 1241--1245.

\bibitem{6887298}
A.~{Yedla}, Y.~{Jian}, P.~S. {Nguyen}, and H.~D. {Pfister}, ``A simple proof of
  {M}axwell saturation for coupled scalar recursions,'' \emph{IEEE Trans. Inf.
  Theory}, vol.~60, no.~11, pp. 6943--6965, 2014.

\bibitem{9174017}
M.~U. Farooq, S.~Moloudi, and M.~Lentmaier, ``Generalized {LDPC} codes with
  convolutional code constraints,'' in \emph{Proc. IEEE Int. Symp. Inf. Theory
  (ISIT)}, 2020, pp. 479--484.

\bibitem{7296605}
C.~{Rachinger}, J.~B. {Huber}, and R.~R. {M\"{u}ller}, ``Comparison of
  convolutional and block codes for low structural delay,'' \emph{IEEE Trans.
  Commun.}, vol.~63, no.~12, pp. 4629--4638, 2015.

\bibitem{7932507}
M.~Zhu, D.~G.~M. Mitchell, M.~Lentmaier, D.~J. Costello, and B.~Bai, ``Braided
  convolutional codes with sliding window decoding,'' \emph{IEEE Trans.
  Commun.}, vol.~65, no.~9, Sept. 2017.

\bibitem{8214245}
R.~Garzón-Bohórquez, C.~Abdel~Nour, and C.~Douillard, ``Protograph-based
  interleavers for punctured turbo codes,'' \emph{IEEE Trans. Commun.},
  vol.~66, no.~5, pp. 1833--1844, 2018.

\bibitem{9594196}
M.~U. Farooq, A.~Graell~i Amat, and M.~Lentmaier, ``Spatially-coupled serially
  concatenated codes with periodic convolutional permutors,'' in \emph{Proc.
  Int. Symp. Turbo Codes Iterative Inf. Process (ISTC)}, 2021, pp. 1--5.

\end{thebibliography}

\end{document}